\documentclass[11pt]{article}

\usepackage[margin=1in]{geometry}
\usepackage{amsmath,amssymb,amsthm}
\usepackage{graphicx}
\usepackage[round,authoryear]{natbib}   
\usepackage{xcolor}
\usepackage{bm}
\usepackage{tikz}
\usetikzlibrary{calc,arrows.meta}
\usepackage{booktabs}
\usepackage[font=small,labelfont=bf]{caption}
\usepackage{newtxtext}
\usepackage{newtxmath}
\usepackage[colorlinks=true,citecolor=blue,linkcolor=blue,urlcolor=blue]{hyperref}

\newenvironment{keywords}
  {\par\medskip\noindent\begingroup\small\textbf{Key words: }}
  {\endgroup\par\medskip}
\newcommand{\lefttitle}[1]{}
\newcommand{\righttitle}[1]{}
\newcommand{\corresau}[1]{}

\definecolor{jfmblue}{RGB}{0,56,168}
\definecolor{cn}{HTML}{C62828}   
\definecolor{ct}{HTML}{D1720A}   
\definecolor{cz}{HTML}{7B1FA2}   

\title{Turbulence at an aerofoil leading edge: the linearity boundary and its generation mechanism}
\author{Sparsh Sharma,\quad Malav Soni,\quad Alexandre Suryadi,\quad Michaela Herr\\[4pt]
  \normalsize German Aerospace Center (DLR), 38108 Braunschweig, Germany\\[2pt]
  \normalsize Corresponding author: Sparsh Sharma, \texttt{sparsh.sharma@dlr.de}}
\date{}

\begin{document}
\maketitle

\begin{abstract}
Turbulence approaching an aerofoil leading edge is strained and blocked
in the final few nose radii. Rapid-distortion theory assumes that the
pre-impact field is a linear image of the upstream turbulence; we test
that assumption directly. Grid turbulence is computed over a NACA~0012
aerofoil and sampled on three congruent surfaces around the nose, giving
the cross-spectral tensor between upstream and pre-impact stations.
Relative to a no-aerofoil control, the linearity spectrum falls by a
factor of three already at the energy-containing scales
($\kappa = k\,r_{\mathrm{LE}} \approx 0.3$), showing that a substantial
part of the pre-impact turbulence is generated locally rather than
inherited. The same location appears in physical space: the anisotropy
invariant $III_b$ changes sign from prolate to oblate within the last
few per cent of chord, departing the plane-strain rapid-distortion
prediction. The change is produced by pressure--strain redistribution,
which drains the strain-amplified wall-normal stress into the other
components; the term is absent in the control and becomes active once
the accumulated strain is rapid. Across a fourteen-fold range of nose
radius the sign change occurs at $x \approx -1.5\,r_{\mathrm{LE}}$ and
the generated turbulence is spanwise-dominated. At second order the
mapping remains usable: when inserted into a standard Amiet prediction
it improves agreement with measured leading-edge noise, and the same
coefficients transfer to independent published cases when only nose
radius and free-stream speed are supplied.
\end{abstract}




\section{Introduction}
\label{sec:introduction}

When turbulence is carried against a bluff body, it is transformed before
it ever reaches the surface. The decelerating mean flow strains each eddy;
the surface blocks the approach of the largest ones; and in the last few
body radii the incident field is reorganised, component by component and
scale by scale. How much of that transformation is captured by
\emph{linear} theory --- each eddy deformed independently, the statistics
following by superposition --- is one of the oldest questions of rapid
distortion theory, posed for bluff bodies by \citet{Hunt1973} and never
fully answered, because answering it requires comparing the same turbulence
before and after the distortion at the level of the field, not merely of
its spectra.

The question has acquired practical weight in aeroacoustics. At low Mach
number the loudest broadband source on an aerofoil is often the leading
edge: incoming eddies load the nose unsteadily and the fluctuating load
radiates. This turbulence-interaction noise governs fans, propellers,
turbomachinery in distorted inflow and, increasingly, wind turbines in the
atmospheric boundary layer
\citep{Amiet1975,PatersonAmiet1976,Moriarty2005,DevenportStaubsGlegg2010,
Yalcin2026}; as trailing-edge and tonal sources have been reduced
it has become an important design consideration
\citep{KimHaeriJoseph2016,SharmaSarradjSchmidt2020,SharmaHerr2024}. And
every standard prediction of it rests, silently, on the linearity question
above.

\begin{figure}
  \centering
  \includegraphics[width=\textwidth]{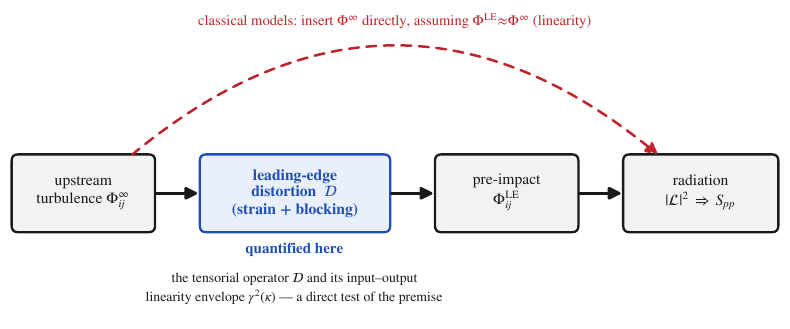}
  \caption{The causal chain of leading-edge noise. Standard models insert the
  upstream spectrum directly into the acoustic transfer function (dashed),
  bypassing the distortion operator $\mathcal{D}$; this work quantifies
  $\mathcal{D}$ and the linearity envelope $\gamma^{2}(\kappa)$ of the bypassed
  step.}
  \label{fig:chain}
\end{figure}

Following \citet{Amiet1975},
the far-field noise spectrum is the product of two factors: an
aerodynamic--acoustic transfer function, which describes how a flat plate
responds to a sinusoidal gust \citep{Sears1941,Graham1970} and radiates
\citep{Lighthill1952,Curle1955,FfowcsWilliamsHawkings1969,Goldstein1976,
Howe2003}, and the spectrum of the incoming turbulence. The turbulence
enters purely as a boundary datum: a spectrum, measured somewhere upstream,
inserted into the formula. This procedure rests on an assumption.
The spectrum that actually forces the leading edge is not the upstream
spectrum --- it is the spectrum of the turbulence \emph{after} it has been
decelerated, strained and squeezed by the flow around the nose. Using the
upstream spectrum assumes that this transformation either does nothing or,
at worst, acts as a known linear filter on each eddy --- the premise of
rapid-distortion theory (RDT). Figure~\ref{fig:chain} draws the causal
chain and marks where the standard models cut it short.

Empirical corrections have been introduced to account for the gap.
High-frequency corrections attenuate the response of a thick section
\citep{Gershfeld2004,RogerMoreau2005}; computations isolated the thickness
effect on the radiated field \citep{Gill2013}; laboratory measurements
separated the roles of thickness and nose radius
\citep{Chaitanya2015,DevenportStaubsGlegg2010}; and corrected upwash
spectra, informed by rapid-distortion theory or by hot-wire surveys near
the nose, restore much of the lost accuracy
\citep{deSantana2016,dosSantos2023}. Semi-analytical models of the
distortion itself are now an active development
\citep{PiccoloJSV2026,Yalcin2026,HalesJFM2023}. Every one of these,
however, operates on a single scalar --- the upwash spectrum. What the
distortion does to the full tensor, and whether it acts linearly at all,
is not addressed by any of them, because no available measurement could
address it.

The elements of the linear theory are classical.
\citet{BatchelorProudman1954} derived the response of weak turbulence to a
rapid mean deformation, and \citet{Hunt1973} specialised it to bluff
bodies, separating two competing effects: the mean strain \emph{amplifies} the velocity
fluctuation normal to the surface, while the surface itself \emph{blocks}
it, with the balance set by the size of an eddy relative to the body
\citep{HuntGraham1978}. Hot-wire measurements upstream of cylinders
confirmed the competition \citep{Bearman1972,BritterHuntMumford1979}, and
the stagnation region has since been studied closely in the heat-transfer
context, where the same distortion controls leading-edge thermal loads
\citep{VanFossen1995,XiongLele2007,WissinkRodi2011}. In aeroacoustics the
picture is actively being rebuilt around this physics: rapid-distortion
corrections to the upwash spectrum improve Amiet-type predictions for thick
aerofoils \citep{deSantana2016,dosSantos2023}, the distortion upstream of an
aerofoil nose has been shown to match that of an equivalent cylinder
\citep{dosSantosAIAA2024}, scale-resolved simulations have traced its
effect on the radiated noise \citep{Piccolo2024}, and distortion has been
invoked to explain the noise reduction of porous leading edges
\citep{Zamponi2020}. All of this work, however, characterises the
transformation in reduced form: a corrected scalar spectrum, a
streamline velocity ratio, a fitted attenuation.

A practical difficulty has been instrumental. A hot wire returns
the spectrum at one point; a pair of probes, the coherence at one
separation; a conventional simulation probe, a time series at a location
chosen in advance. The object the linearity question requires is richer
than any of these: the full cross-spectral \emph{tensor} between the
turbulence upstream and the turbulence about to strike the nose, taken at
matched spatial positions and on a common clock, so that the two states of
the same field can be compared rather than merely their statistics. No
laboratory or numerical data set has provided it, and in its absence three
questions of principle have remained open:
\begin{enumerate}
  \item Is the mapping from the upstream turbulence tensor
    $\Phi^{\infty}_{ij}$ to the pre-impact tensor
    $\Phi^{\mathrm{LE}}_{ij}$ a scalar, frequency-dependent gain --- or is
    it irreducibly tensorial, treating different velocity components
    differently?
  \item Is its scale dependence organised by the leading-edge radius
    $r_{\mathrm{LE}}$, as far-field acoustic trends suggest
    \citep{Gill2013,DevenportStaubsGlegg2010} but no direct evaluation on
    the turbulence field has shown?
  \item And --- the question beneath both --- how far is the
    transformation \emph{linear} at all?
\end{enumerate}
Without answers to these questions, the error incurred by inserting an upstream spectrum into a leading-edge noise model remains unquantified.

We address the three questions with a computation designed for the purpose. We
simulate grid turbulence convecting onto a NACA~0012 aerofoil, reproducing a
laboratory experiment, and sample the turbulence on three geometrically
congruent surfaces wrapped around the leading edge --- an arrangement whose
vertex-for-vertex correspondence yields the full cross-spectral tensor
between the upstream and pre-impact states, something neither a hot wire
nor a conventional simulation probe can provide. A preliminary account was
given in \citet{Sharma2026}.

The straining and blocking of turbulence at a bluff nose are classical
\citep{Hunt1973,HuntGraham1978}, the two-component state at the surface is
their anticipated limit, and the growth of the lateral scales near a
leading edge has been measured before \citep{dosSantos2023}. The present
work rests on three congruent sampling surfaces
(figure~\ref{fig:instrument}) that allow four quantities to be obtained:
\begin{itemize}
  \item a scale-resolved determination of the \emph{linearity} of the
    distortion, the spectrum $\gamma^{2}(\kappa)$, with its decomposition
    of the pre-impact tensor into inherited and generated parts;
  \item the localisation of the anisotropy-type inversion, and its
    collapse on the nose radius across a fourteen-fold geometry sweep;
  \item a surface-based evaluation of the pressure--strain tensor at the
    stagnation line, tied causally to that inversion;
  \item the sector-antisymmetric, off-diagonal structure of the generated
    turbulence.
\end{itemize}
The remaining results (tensorial gains, coherence stretching and the acoustic correction) provide incremental detail on existing knowledge and are presented accordingly.

The main findings fall into two parts, and the paper is organised around them.
Section~\ref{sec:numerical} describes the flow, the laboratory reference
and the sampling instrument --- and shows what the turbulence approaching
the nose actually looks like. Section~\ref{sec:results} addresses the linearity
question directly. The distortion proves strong, organised by the reduced wavenumber
$\kappa = k\,r_{\mathrm{LE}}$, and tensorial; but its
\emph{linearity} collapses at the largest, energy-containing scales, and
the anisotropy of the turbulence changes type within the final few per
cent of chord --- most of the turbulence that strikes the nose is generated
on the way in, not inherited from upstream. Section~\ref{sec:mechanism}
identifies the dynamics responsible: pressure--strain redistribution,
evaluated on the stagnation line by a route that makes an ordinarily
inaccessible quantity computable on a single surface, and shown to be
robust across a fourteen-fold sweep of nose radius.
Section~\ref{sec:results_acoustic} quantifies the acoustic consequence
against an independent laboratory experiment, \S\ref{sec:generality} discusses what
organises the scales, and \S\ref{sec:conclusions} concludes. 

\section{Flow configuration, experiment, and sampling}
\label{sec:numerical}
\label{sec:framework_states}
\label{sec:method_configuration}

The requirement identified in \S\ref{sec:introduction} --- the same
turbulence, characterised before and after distortion at the level of the
field --- dictated every choice described below. The strategy is to
reproduce a laboratory flow closely enough that it can then be
interrogated in ways no laboratory allows: this section presents the flow,
the wind-tunnel experiment that anchors it, and the sampling arrangement
built to meet that requirement.

\subsection{Configuration and governing parameters}
\label{sec:framework_groups}

Grid-generated turbulence convects onto a NACA~0012 aerofoil of chord
$c=0.400$~m at zero incidence, in a uniform stream
$U_\infty=20$~m\,s$^{-1}$ ($Re_c\approx5.4\times10^{5}$, $M=0.058$). The
turbulence is produced by an explicitly resolved passive grid --- square
bars of width $d_g=20$~mm at pitch $M_g=60$~mm, solidity
$\sigma\approx0.56$ --- placed $2.19\,c$ upstream of the nose, replicating
the hardware of the DLR Acoustic Wind Tunnel Braunschweig (AWB) described
below. The grid-to-aerofoil distance ($\approx14.6\,M_g$) is long enough for
the bar wakes to merge into a developed broadband field
\citep{ComteBellotCorrsin1966,Roach1987}, short enough that the
energy-containing eddies arrive still energetic. We resolve the grid rather
than prescribe a synthetic inflow for one reason: a synthetic method fixes
the statistical state of the turbulence by construction, and that state is
precisely what is under investigation here.

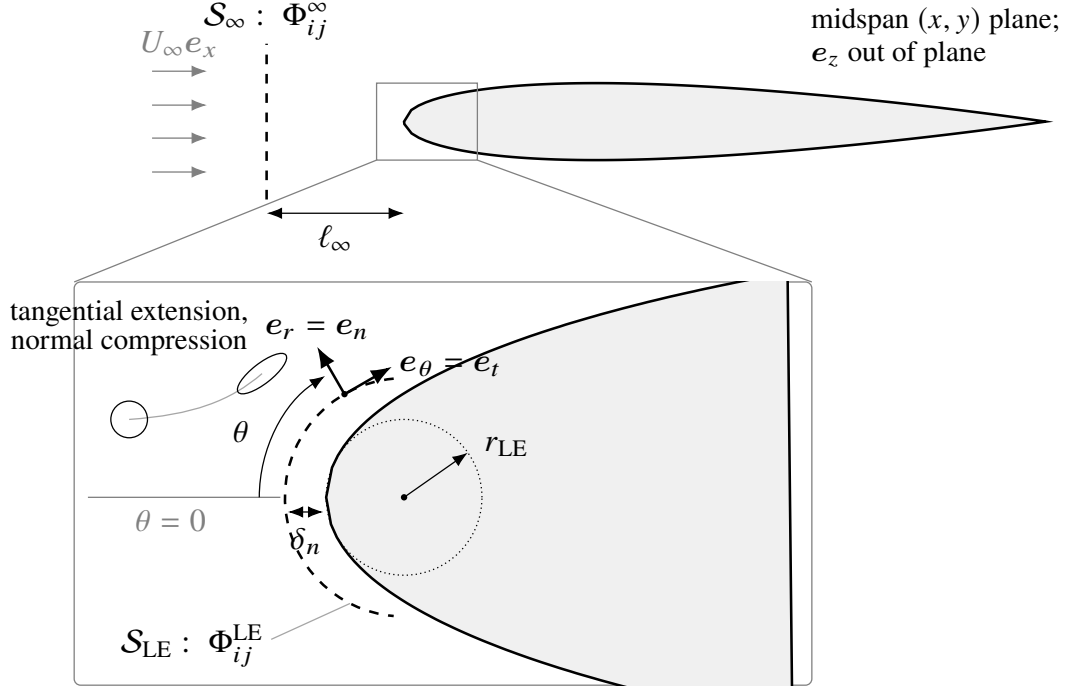
\begin{figure}
\centering
\resizebox{0.86\textwidth}{!}{%
\begin{tikzpicture}[>=Latex, font=\small]
\begin{scope}[shift={(0,4.1)}]
  \foreach \y in {-0.55,-0.18,0.18,0.55}
    \draw[->,gray] (-2.75,\y) -- (-2.15,\y);
  \node[gray] at (-2.45,0.85) {$U_\infty\boldsymbol{e}_x$};
  \draw[dashed,thick] (-1.5,-0.85) -- (-1.5,0.85);
  \node at (-1.5,1.10) {$\mathcal{S}_\infty:\;\Phi^{\infty}_{ij}$};
  \draw[<->] (-1.5,-1.00) -- (0,-1.00);
  \node[below] at (-0.75,-1.00) {$\ell_\infty$};
  \draw[thick,fill=gray!12]
    plot[domain=0:1,samples=90]
      ({7*\x},{4.2*(0.2969*sqrt(\x)-0.1260*\x-0.3516*\x*\x
               +0.2843*\x*\x*\x-0.1036*\x*\x*\x*\x)})
    -- plot[domain=1:0,samples=90]
      ({7*\x},{-4.2*(0.2969*sqrt(\x)-0.1260*\x-0.3516*\x*\x
               +0.2843*\x*\x*\x-0.1036*\x*\x*\x*\x)})
    -- cycle;
  \node[align=left] at (5.8,0.92)
        {\footnotesize midspan $(x,y)$ plane;\\[-2pt]
         \footnotesize $\boldsymbol{e}_z$ out of plane};
  \draw[gray] (-0.30,-0.42) rectangle (0.80,0.42);
\end{scope}
\draw[gray,thin] (-0.30,3.68) -- (-3.60,2.35);
\draw[gray,thin] (0.80,3.68) -- (4.45,2.35);
\draw[gray,rounded corners=2pt] (-3.60,-2.05) rectangle (4.45,2.35);
\begin{scope}
  \clip (-3.60,-2.05) rectangle (4.45,2.35);
  \draw[thick,fill=gray!12]
    plot[domain=0:0.095,samples=80]
      ({53.55*\x-0.85},{32.13*(0.2969*sqrt(\x)-0.1260*\x-0.3516*\x*\x
               +0.2843*\x*\x*\x-0.1036*\x*\x*\x*\x)})
    -- plot[domain=0.095:0,samples=80]
      ({53.55*\x-0.85},{-32.13*(0.2969*sqrt(\x)-0.1260*\x-0.3516*\x*\x
               +0.2843*\x*\x*\x-0.1036*\x*\x*\x*\x)})
    -- cycle;
\end{scope}
\draw[densely dotted] (0,0) circle (0.85);
\node[circle,fill,inner sep=0.7pt] at (0,0) {};
\draw[->] (0,0) -- (35:0.85);
\node at (1.12,0.55) {$r_{\mathrm{LE}}$};
\draw[dashed,thick] (95:1.30) arc (95:265:1.30);
\draw[gray!70] (243:1.30) -- (-1.45,-1.55);
\node[anchor=east] at (-1.40,-1.62) {$\mathcal{S}_{\mathrm{LE}}:\;\Phi^{\mathrm{LE}}_{ij}$};
\draw[<->] (-0.88,-0.16) -- (-1.27,-0.16);
\node[below] at (-1.07,-0.20) {$\delta_n$};
\draw[gray] (-3.45,0) -- (-1.35,0);
\node[gray,below] at (-2.55,0) {$\theta=0$};
\draw[->] (-1.58,0) arc (180:123:1.58);
\node at (-1.78,0.72) {$\theta$};
\coordinate (P) at (120:1.30);
\node[circle,fill,inner sep=0.7pt] at (P) {};
\draw[->,thick] (P) -- ++(120:0.60);
\node[anchor=south,inner sep=2pt] at ($(P)+(120:0.60)$)
      {$\boldsymbol{e}_r=\boldsymbol{e}_n$};
\draw[->,thick] (P) -- ++(30:0.60);
\node[anchor=west,inner sep=2pt] at ($(P)+(30:0.60)$)
      {$\boldsymbol{e}_\theta=\boldsymbol{e}_t$};
\draw[gray!70] (-3.0,0.85) .. controls (-2.3,0.90) and (-1.9,1.05) .. (-1.55,1.35);
\draw (-3.0,0.85) circle (0.20);
\draw[rotate around={38:(-1.55,1.35)}] (-1.55,1.35) ellipse (0.33 and 0.11);
\node[align=center] at (-3.,1.85)
      {\footnotesize tangential extension,\\[-3pt]\footnotesize normal compression};
\end{tikzpicture}}
\caption{Configuration and definitions. Top: upstream reference plane
  $\mathcal{S}_\infty$ ahead of the aerofoil. Bottom: magnified nose with the
  osculating circle (radius $r_{\mathrm{LE}}$), the conformal sampling surface
  $\mathcal{S}_{\mathrm{LE}}$ at offset $\delta_n \ll r_{\mathrm{LE}}$, and the
  azimuth $\theta$ measured from the stagnation line, where
  $\boldsymbol{e}_r=\boldsymbol{e}_n$ and
  $\boldsymbol{e}_\theta=\boldsymbol{e}_t$.}
\label{fig:framework_config}
\end{figure}

Two numbers organise everything in this paper, and both follow from the
geometry. The nose of a NACA~0012 is locally a cylinder of radius
$r_{\mathrm{LE}}=1.1019\,t_c^{2}\,c\approx6.3$~mm
\citep{AbbottDoenhoff1959}. The first number compares an eddy to that
nose,
\begin{equation}
  \kappa = k\,r_{\mathrm{LE}},
  \label{eq:kappa_def}
\end{equation}
with $k$ the eddy wavenumber: $\kappa\ll1$ is an eddy much larger than the
nose, $\kappa\gg1$ one much smaller. The second measures how much straining
an eddy accumulates on its way in. An ideal stagnation flow about the nose
has strain rate
\begin{equation}
  a \sim \frac{U_{\infty}}{2\,r_{\mathrm{LE}}}
    \approx 1.6\times10^{3}~\mathrm{s}^{-1},
  \label{eq:strain_scale}
\end{equation}
and a residence time $\tau\sim r_{\mathrm{LE}}/U_\infty$ in the strained
region, so the integrated distortion is a priori
\begin{equation}
  \vartheta = a\tau = \mathcal{O}(1/2).
  \label{eq:vartheta_estimate}
\end{equation}
These are geometric estimates; both are computed from the data later
(\S\ref{sec:results_volume}), where $\vartheta\approx0.68$ on the
stagnation line. The configuration, the sampling surfaces and the local
frames are summarised in figure~\ref{fig:framework_config}: points near the
nose are parametrised by the azimuth $\theta$ about the osculating nose
circle, with $\theta=0$ on the upstream stagnation line, where the
surface-normal direction $n$ coincides with the streamwise direction and
the tangential direction $t$ with the transverse one; $z$ is spanwise.

\subsection{The computation}

The flow is computed by wall-modelled large-eddy simulation with the
lattice Boltzmann solver
ProLB~3.4.3 \citep{ProLB}, using the hybrid recursive regularised collision
operator of \citet{Jacob2018}, the shear-improved Smagorinsky subgrid model
\citep{Leveque2007} and a wall-function treatment at solid surfaces
\citep{Wilhelm2018}; refinement levels are coupled following
\citet{Touil2014}. The wall treatment carries little weight for what
follows: every quantity in this paper is evaluated \emph{upstream} of the
boundary layer, in the decelerating approach flow, and the pre-impact
statistics are mesh-converged (supplementary material). Two properties of
the method do matter: low intrinsic
dissipation and dispersion, which carry the grid turbulence over the
$2.19\,c$ approach without significant numerical damping, and a
compressible time-domain formulation, which keeps the acoustic degrees of
freedom in the solution. The lattice is block-refined over eight levels,
from $\Delta x_{\min}=0.5$~mm at the aerofoil surface
($r_{\mathrm{LE}}/\Delta x_{\min}=12.6$) to $128$~mm in the far field, with
the entire grid-to-aerofoil corridor at $4$~mm --- about ten cells per
wavelength at the crossover scale $\kappa=1$, and about eighty in the
near-nose region where the distortion is interrogated. The domain
(schematic in the supplementary material) extends $\pm$ tens of chords with
absorbing layers on all outer boundaries; the aerofoil spans the full
$L_z=0.800$~m between free-slip walls. The mesh comprises
$1.10\times10^{9}$ fluid nodes. The simulation runs for $33.3$ flow-through
times; the first half is discarded and all statistics accumulate over the
remaining $0.333$~s, whose stationarity is verified by half-record
comparison. Table~\ref{tab:config} collects the numbers.

\begin{table}
  \begin{center}
  \def~{\hphantom{0}}
  \begin{tabular}{lll}
    Quantity & Symbol & Value \\[3pt]
    Aerofoil section                  & ---                & NACA~0012, zero incidence \\
    Chord                            & $c$                & $0.400$~m \\
    Span (slip-wall to slip-wall)    & $L_z$              & $0.800$~m \quad ($L_z/c = 2$) \\
    Leading-edge radius              & $r_{\mathrm{LE}}$  & $6.3$~mm \quad ($r_{\mathrm{LE}}/c = 1.6\times10^{-2}$) \\
    Grid--leading-edge separation    & $x_{\mathrm{LE}}$  & $0.875$~m \quad ($2.19\,c = 14.6\,M_g$) \\
    Grid bar width / mesh size       & $d_g$ / $M_g$      & $20$~mm / $60$~mm \quad ($\sigma \approx 0.56$) \\
    Free-stream velocity             & $U_{\infty}$       & $20$~m\,s$^{-1}$ \\
    Mach number                      & $M$                & $0.058$ \\
    Chord Reynolds number            & $Re_c$             & $5.4\times10^{5}$ \\
    Stagnation strain-rate scale     & $a$                & $\approx 1.6\times10^{3}$~s$^{-1}$ (Eq.~\ref{eq:strain_scale}) \\
    Distortion parameter             & $\vartheta$        & $\mathcal{O}(1/2)$ est.\ (Eq.~\ref{eq:vartheta_estimate}); $0.68$ computed \\
    Finest lattice spacing           & $\Delta x_{\min}$  & $0.5$~mm \quad ($r_{\mathrm{LE}}/\Delta x_{\min} = 12.6$) \\
    Grid-corridor lattice spacing    & ---                & $4$~mm \quad ($\approx 10$ cells per $\lambda(\kappa{=}1)$) \\
    Refinement levels                & ---                & $8$ \quad ($\Delta x$: $0.5$--$128$~mm) \\
    Fluid nodes                      & ---                & $1.10\times10^{9}$ \\
    Fine time step                   & $\Delta t$         & $8.41\times10^{-7}$~s \\
    Total run / retained record      & ---                & $33.3$ / $16.7$ $\times\,c/U_{\infty}$ \\
    Surface sampling rate            & $f_s$              & $74.3$~kHz \quad (Nyquist $37.2$~kHz) \\
    Surface record                   & ---                & $24\,769$ snapshots over $0.333$~s \\
    Surface points / cells (each)    & ---                & $181\,373$ / $360\,696$ \\
  \end{tabular}
  \caption{Summary of the computational configuration (problem
  definition for the simulation in ProLB~3.4.3).}
  \label{tab:config}
  \end{center}
\end{table}

\subsection{Sampling: three congruent surfaces}

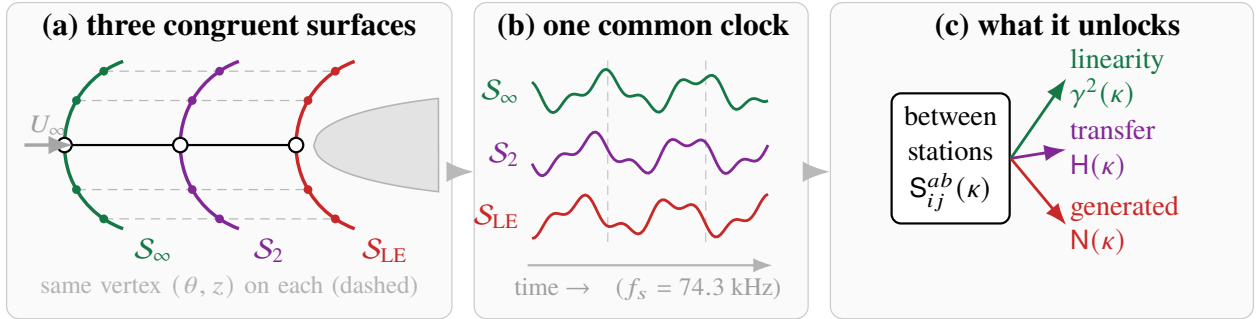
\begin{figure}
\centering
\resizebox{\textwidth}{!}{%
\begin{tikzpicture}[>=Latex, font=\small, every node/.style={inner sep=1pt}]
\definecolor{cSi}{RGB}{20,120,70}   
\definecolor{cS2}{RGB}{130,40,150}  
\definecolor{cSL}{RGB}{200,40,40}   
\draw[rounded corners=4pt, gray!40, fill=gray!4] (-0.45,-1.95) rectangle (4.55,1.60);
\draw[rounded corners=4pt, gray!40, fill=gray!4] ( 4.80,-1.95) rectangle (8.55,1.60);
\draw[rounded corners=4pt, gray!40, fill=gray!4] ( 8.80,-1.95) rectangle (13.55,1.60);
\draw[->, gray!55, line width=1.3pt] (4.55,-0.30) -- (4.80,-0.30);
\draw[->, gray!55, line width=1.3pt] (8.55,-0.30) -- (8.80,-0.30);
\begin{scope}
  \node[anchor=north, font=\small\bfseries] at (2.05,1.52) {(a)~three congruent surfaces};
  \foreach \cx/\cc in {1.2/cSi, 2.5/cS2, 3.8/cSL}{
    \draw[\cc, line width=1.1pt] ([shift={(\cx,0)}]110:1.0) arc (110:250:1.0);
  }
  \fill[gray!22, draw=gray!60, line width=0.5pt]
    plot[domain=0:0.14, samples=40]
      ({3.0+10*\x},{ 10*0.6*(0.2969*sqrt(\x)-0.1260*\x-0.3516*\x*\x+0.2843*\x*\x*\x)})
    -- plot[domain=0.14:0, samples=40]
      ({3.0+10*\x},{-10*0.6*(0.2969*sqrt(\x)-0.1260*\x-0.3516*\x*\x+0.2843*\x*\x*\x)})
    -- cycle;
  \foreach \a in {124,150,210,236}{
    \draw[gray!55, densely dashed, line width=0.4pt]
      ([shift={(1.2,0)}]\a:1.0) -- ([shift={(2.5,0)}]\a:1.0) -- ([shift={(3.8,0)}]\a:1.0);
    \foreach \cx/\cc in {1.2/cSi,2.5/cS2,3.8/cSL}{
      \fill[\cc] ([shift={(\cx,0)}]\a:1.0) circle (1.4pt);}
  }
  \draw[black, line width=0.7pt] (0.2,0) -- (2.8,0);
  \foreach \cx in {1.2,2.5,3.8}{\draw[black, line width=0.7pt, fill=white] ([shift={(\cx,0)}]180:1.0) circle (2.3pt);}
  \draw[->, gray!70, line width=1.1pt] (-0.25,0) -- (0.30,0);
  \node[gray!75, anchor=south, font=\scriptsize] at (0.02,0.06) {$U_\infty$};
  \node[cSi, anchor=north, font=\footnotesize] at (1.2,-1.00) {$\mathcal{S}_\infty$};
  \node[cS2, anchor=north, font=\footnotesize] at (2.5,-1.00) {$\mathcal{S}_2$};
  \node[cSL, anchor=north, font=\footnotesize] at (3.8,-1.00) {$\mathcal{S}_{\mathrm{LE}}$};
  \node[gray!60, anchor=north, font=\scriptsize, text width=4.6cm, align=center] at (2.05,-1.42)
    {same vertex $(\theta,z)$ on each (dashed)};
\end{scope}
\begin{scope}[shift={(4.95,0)}]
  \node[anchor=north, font=\small\bfseries] at (1.75,1.52) {(b)~one common clock};
  \foreach \tx in {1.35,2.45}{\draw[gray!45, densely dashed, line width=0.4pt] (\tx,0.95) -- (\tx,-1.10);}
  \foreach \yy/\cc/\ph/\lab in {%
      0.60/cSi/0/{\mathcal{S}_\infty}, -0.10/cS2/50/{\mathcal{S}_2}, -0.80/cSL/95/{\mathcal{S}_{\mathrm{LE}}}}{
    \node[\cc, anchor=east, font=\footnotesize] at (0.40,\yy) {$\lab$};
    \draw[\cc, line width=0.9pt] plot[domain=0.50:3.15, samples=90]
      ({\x},{\yy+0.17*sin(340*\x+\ph)+0.08*sin(880*\x+2*\ph)});}
  \draw[->, gray!60, line width=0.7pt] (0.45,-1.35) -- (3.20,-1.35);
  \node[gray!75, anchor=north, font=\scriptsize] at (1.80,-1.45) {time~$\to$\quad($f_s=74.3$~kHz)};
\end{scope}
\begin{scope}[shift={(9.30,0)}]
  \node[anchor=north, font=\small\bfseries] at (2.05,1.52) {(c)~what it unlocks};
  \node[draw, rounded corners=3pt, fill=white, line width=0.7pt, align=center,
        font=\footnotesize, inner sep=4pt] (S) at (0.85,-0.15)
    {between\\stations\\[1pt]$\mathsf{S}^{ab}_{ij}(\kappa)$};
  \node[anchor=west, cSi, font=\footnotesize, text width=1.9cm] (g) at (2.15,0.75)
    {linearity\\ $\gamma^{2}(\kappa)$};
  \node[anchor=west, cS2, font=\footnotesize, text width=1.9cm] (h) at (2.15,-0.05)
    {transfer\\ $\mathsf{H}(\kappa)$};
  \node[anchor=west, cSL, font=\footnotesize, text width=1.9cm] (n) at (2.15,-0.90)
    {generated\\ $\mathsf{N}(\kappa)$};
  \draw[->, cSi, line width=0.9pt] (S.east) -- (g.west);
  \draw[->, cS2, line width=0.9pt] (S.east) -- (h.west);
  \draw[->, cSL, line width=0.9pt] (S.east) -- (n.west);
\end{scope}
\end{tikzpicture}%
}
\caption{The sampling principle. (\textit{a})~Three geometrically
congruent surfaces --- upstream $\mathcal{S}_\infty$, midpoint
$\mathcal{S}_2$, pre-impact $\mathcal{S}_{\mathrm{LE}}$ --- place every
vertex $(\theta,z)$ in three-station correspondence (dashed).
(\textit{b})~All three are sampled on one common clock, so the records are
simultaneous. (\textit{c})~The between-station cross-spectral tensor
$\mathsf{S}^{ab}_{ij}(\kappa)$ then yields the linearity spectrum
$\gamma^{2}$, the transfer $\mathsf{H}$ and the generated tensor
$\mathsf{N}$ --- statistics that require matched stations on a common clock.}
\label{fig:instrument}
\end{figure}

The analysis rests on one feature of the sampling
(figure~\ref{fig:instrument}).
Three concentric, geometrically congruent toroidal surfaces of identical
topology ($181\,373$ points each) wrap the leading edge across the full
span and are recorded synchronously at $74.3$~kHz --- $24\,769$ snapshots
of velocity and pressure. The innermost, $\mathcal{S}_{\mathrm{LE}}$,
conforms to the nose at ${\approx}1\,r_{\mathrm{LE}}$ and carries the
\emph{pre-impact} state; the outermost, $\mathcal{S}_\infty$, sits
$51\,r_{\mathrm{LE}}$ ($0.80\,c$) upstream and carries the \emph{upstream}
state; $\mathcal{S}_2$ lies midway. Because the surfaces are congruent and
share a common time base, every vertex on one corresponds to exactly one
vertex on another, through a fixed $(\theta,z)$ relabelling. It is this
correspondence that lets us ask, later, not just \emph{how much} turbulence
arrives at the nose but \emph{whether it is the same turbulence} that left
the upstream station. Two further record families support the analysis: a
midspan probe line of $200$ stations for mean profiles and Reynolds
stresses along the approach, and a ladder of $13$ cross-flow planes
carrying the mean Reynolds tensor to within $10^{-2}c$ of the nose.

Spectral estimation is standard and stated once.
\label{sec:methodology_gain}
All spectra are formed by Welch averaging (Hann window, $50\%$ overlap;
the estimator verified against a reference implementation to
${\sim}10^{-7}$), averaged over segments and over the vertices of a
poloidal sector (a patch of the surface covering a range of azimuth
$\theta$); $n_{\mathrm{seg}}=2048$ serves the converged
mid-$\kappa$ estimates and $n_{\mathrm{seg}}=4096$ the largest scales.
Between-station quantities are read in a common frame: the vertex
correspondence is exact to machine precision, and the bulk convective lag
between stations --- which would otherwise destroy any coherence --- is
removed by reading each station at the matching integer sample offset. To
place a frequency-domain spectrum on the eddy-size axis $\kappa$ we use
frozen convection \citep{Taylor1938},
\begin{equation}
  \kappa(f) = \frac{2\pi f\,r_{\mathrm{LE}}}{U_c(f)},
  \label{eq:kappa_f}
\end{equation}
with the convection velocity $U_c$ taken from computed cross-spectral
phases on the upstream side, where frozen convection is secure (checks in
the supplementary material); no conclusion below rests on the precise
location of the $\kappa=\mathcal{O}(1)$ crossover.

\subsection{The laboratory reference}
\label{sec:ExpSetup}

The laboratory data were acquired in the AWB, a closed-circuit, anechoic,
open test-section facility \citep{PottPollenskeDelfs2008} (anechoic above
${\approx}200$~Hz; residual turbulence intensity ${\approx}0.3\%$; nozzle
$800\times1200$~mm), with the same passive grid mounted $495$~mm upstream
of the nozzle exit; figure~\ref{fig:awb} shows the installation, and
further photographs are collected in the supplementary material.

\begin{figure}
  \centering
  \begin{minipage}{0.585\textwidth}
    \includegraphics[width=\textwidth]{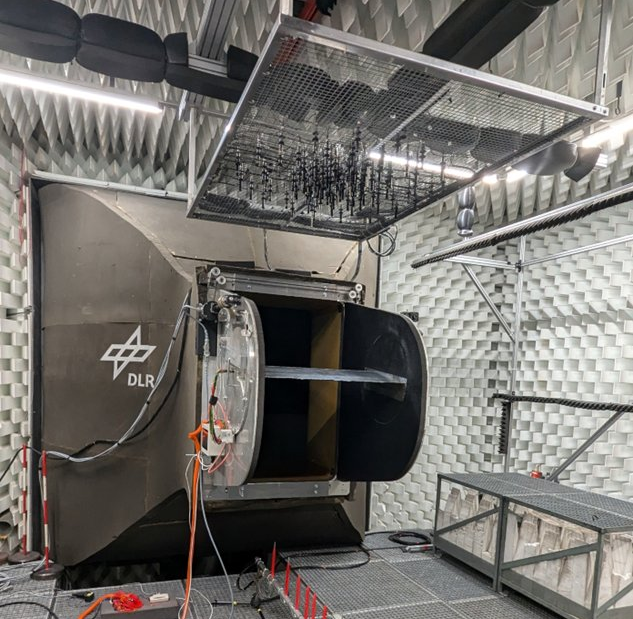}
  \end{minipage}\hfill
  \begin{minipage}{0.395\textwidth}
    \includegraphics[width=\textwidth]{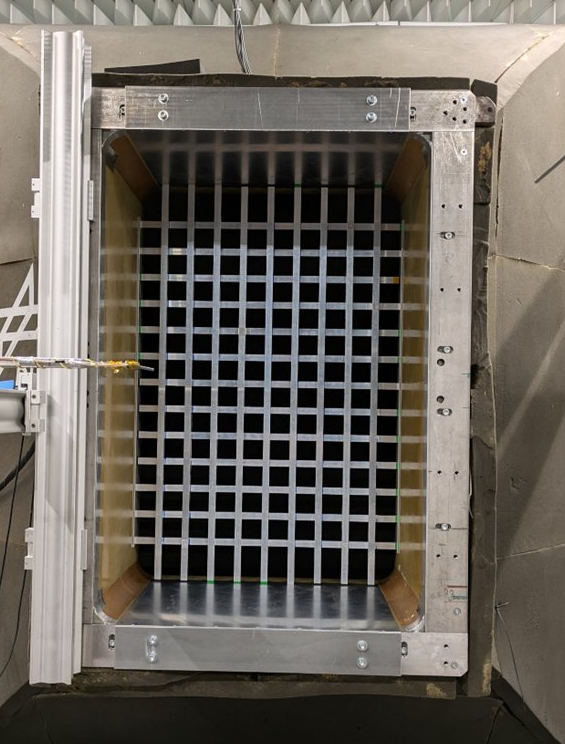}
  \end{minipage}
  \caption{The laboratory reference in the DLR Acoustic Wind Tunnel
  Braunschweig: (left)~open test section with the aerofoil installed;
  (right)~the turbulence grid at the nozzle.}
  \label{fig:awb}
\end{figure}
\emph{Velocity}: two Dantec~55P51 cross-wire probes ($5\,\mu$m wires) sat
$380$~mm downstream of the exit --- the same grid-to-probe distance as the
computation's grid-to-leading-edge distance, so simulated and measured
inflow can be compared at matched development length. One probe traversed
the midspan, the second sat $200$~mm off-midspan for simultaneous two-point
measurements; the third velocity component was obtained by rotating the
probe $90^\circ$ in separate runs; calibration details are given in the
supplementary material. \emph{Acoustics}: grid-generated inflow raises the
background noise, so the isolated leading-edge noise was extracted by the
coherent-output-power (COP) method from two far-field microphones placed at
$\pm\varphi$ about the chord plane --- edge noise is dipole-like, so the
two signals are in anti-phase and their cross-spectrum estimates the dipole
power while incoherent background cancels. That the dominant source is the
leading edge was confirmed from reception times at streamwise-separated
microphones. No formal uncertainty is assigned to the COP spectrum, and
acoustic comparisons below are quoted as mean absolute deviations.

In addition to the primary computation (case A0), three companions close
the loop, listed in table~\ref{tab:matrix}. A no-aerofoil \emph{control} ---
identical domain, grid, resolution and sampling, aerofoil removed --- lets
every result be read as ``aerofoil versus no aerofoil'' rather than
``downstream versus upstream''. And a nose-radius \emph{sweep}, three
further sections spanning with A0 a fourteen-fold range of
$r_{\mathrm{LE}}$, each an otherwise identical
${\sim}1.1\times10^{9}$-node computation, tests later whether anything
found here is special to one geometry.

\begin{table}
  \begin{center}
  \def~{\hphantom{0}}
  \begin{tabular}{llccl p{4.6cm}}
    Case & Geometry & Grid & $U_{\infty}$ (m\,s$^{-1}$) & $r_{\mathrm{LE}}$ (mm) & Role \\[3pt]
    A0 & NACA 0012        & yes & 20 & $6.3$  & primary case: full operator identification (\S\S\ref{sec:results},~\ref{sec:methodology_gain}) \\
    B0 & none (grid only) & yes & 20 & ---    & no-aerofoil control: free-decay reference and aerofoil-imprint isolation (\S\ref{sec:results_control}) \\
    S1 & NACA 66-006      & yes & 20 & $0.8$  & nose-radius sweep (\S\ref{sec:results_sweep}) \\
    S2 & NACA 0008        & yes & 20 & $2.8$  & nose-radius sweep (\S\ref{sec:results_sweep}) \\
    S3 & NACA 0016        & yes & 20 & $11.2$ & nose-radius sweep (\S\ref{sec:results_sweep}) \\
  \end{tabular}
  \caption{Simulation matrix: aerofoil case A0, geometrically identical
  no-aerofoil control B0 (configuration of table~\ref{tab:config}), and sweep
  cases S1--S3 with the aerofoil section exchanged;
  $Re_c=5.4\times10^{5}$, $M=0.058$ throughout.}
  \label{tab:matrix}
  \end{center}
\end{table}

\subsection{The incident turbulence}
\label{sec:results_incident}

\begin{figure}
  \centering
  \includegraphics[width=0.82\textwidth]{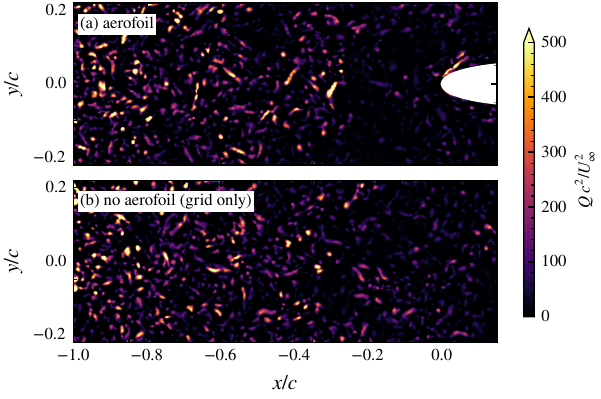}
  \caption{$Q$-criterion $Q\,c^2/U_\infty^2$ on the midspan plane (single
  realisation): (\textit{a})~aerofoil, (\textit{b})~control. Note the
  vortex-depleted zone at the stagnation region in (\textit{a}). An interactive \href{https://cocalc.ai/share/tCMkBMv2Psqe}{JFM Notebook} is available.}
  \label{fig:qcrit}
\end{figure}

\begin{figure}
  \centering
  \includegraphics[width=\textwidth]{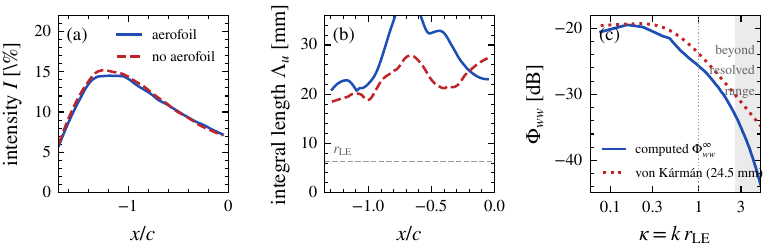}
  \caption{The incident turbulence: (\textit{a})~streamwise intensity and
  (\textit{b})~integral length along the midspan approach, aerofoil versus
  control; (\textit{c})~upstream transverse spectrum against the von
  K\'arm\'an form (shading: beyond the resolved range). An interactive \href{https://cocalc.ai/share/bJyiEi5dXjJj}{JFM Notebook} is available.}
  \label{fig:incident}
\end{figure}

Figure~\ref{fig:qcrit} shows the instantaneous flow: incident grid
turbulence, resolved down to its vortex cores, convecting onto the leading
edge. Even a single snapshot carries a hint of the physics to come --- a
vortex-depleted zone forms at the immediate stagnation region with the
aerofoil and is absent in the control, the kinematic footprint of the
blocking. Figure~\ref{fig:incident} characterises the incident state
quantitatively. In its developed region the turbulence has streamwise
intensity $I\approx8$--$12\%$ and integral length
$\Lambda_u\approx25$~mm~$\approx4\,r_{\mathrm{LE}}$
(figure~\ref{fig:incident}\textit{a},\textit{b}), so its
energy-containing eddies sit near $\kappa\approx0.3$ --- a value that
recurs throughout the analysis. The upstream spectrum follows the von
K\'arm\'an form through the energy-containing range
(figure~\ref{fig:incident}\textit{c}; transverse integral scale
$\Lambda\approx24.5$~mm, consistent with the hot-wire measurement), and
all three stations stay spectrally resolved to $\kappa\approx2$--$3$
(shaded in figure~\ref{fig:incident}\textit{c}; mesh-independence check
in the supplementary material); every claim below is confined to
$\kappa\lesssim2$.

A statistical statement, however, needs more than a snapshot. With the
sampling in place and the incident state characterised, the question posed
in \S\ref{sec:introduction} can be asked of the field itself: when the
turbulence arrives at the nose, how much of it is still, in any linear
sense, the turbulence that left the upstream station?

\section{How linear is the distortion?}
\label{sec:results}
\label{sec:framework_hypothesis}

This section addresses that question in three passes of increasing
depth: first the \emph{amplitude} of the distortion, then its
\emph{linearity}, then its imprint in \emph{physical space}. To make it
precise we write the transformation as a mapping between spectral
tensors,
\begin{equation}
  \Phi^{\mathrm{LE}}_{ij}(\boldsymbol{k},\omega;\theta)
  = \mathcal{D}_{ijmn}(\boldsymbol{k},\vartheta;\theta)\,
    \Phi^{\infty}_{mn}(\boldsymbol{k},\omega),
  \label{eq:operator}
\end{equation}
where $\Phi_{ij}$ is the (Hermitian, positive semi-definite) cross-spectral
tensor of the velocity fluctuations, evaluated upstream ($\infty$) and just
ahead of the nose ($\mathrm{LE}$), and $\mathcal{D}$ is the distortion
operator we wish to characterise --- a diagnostic frame, not a claimed
closure. Any physically admissible $\mathcal{D}$ must map realisable
spectra to realisable spectra --- automatic when it acts by
\emph{congruence}, the sandwich form
$\Phi^{\mathrm{LE}}=\mathsf{M}\,\Phi^{\infty}\mathsf{M}^{\dagger}$ that any
linear transformation of the velocity field induces on its spectrum --- and
must reduce to the identity when no strain has accumulated. Linear
rapid-distortion theory supplies exactly such a congruence
\citep{BatchelorProudman1954,Hunt1973}; whether the flow obeys it is the
question.

\subsection{A strong, tensorial distortion}
\label{sec:results_tensorial}

The first, simplest diagnostic is a ratio of like-component power spectra
between the two stations,
\begin{equation}
  G_q(f)
  = \frac{\Phi^{\mathrm{LE}}_{qq}(f)}{\Phi^{\infty}_{qq}(f)},
  \qquad q \in \{r,\theta,z\}
  \quad\text{(no summation)},
  \label{eq:gain}
\end{equation}
one gain per velocity component, with the in-plane components projected
onto the local cylindrical frame of figure~\ref{fig:framework_config},
\begin{equation}
  u_r = u\cos\theta + v\sin\theta,
  \qquad
  u_\theta = -\,u\sin\theta + v\cos\theta,
  \qquad
  u_z = w.
  \label{eq:projection}
\end{equation}
This rotation preserves the in-plane trace
($\Phi_{uu}+\Phi_{vv}=\Phi_{rr}+\Phi_{\theta\theta}$), so any inequality
between the $r$ and $\theta$ gains is genuine directional redistribution,
not an artefact of the frame. If the distortion were a scalar filter --- one
number per frequency, as a corrected upwash spectrum implicitly assumes ---
the three gains would coincide.

\begin{figure}
  \centering
  \includegraphics[width=0.9\textwidth]{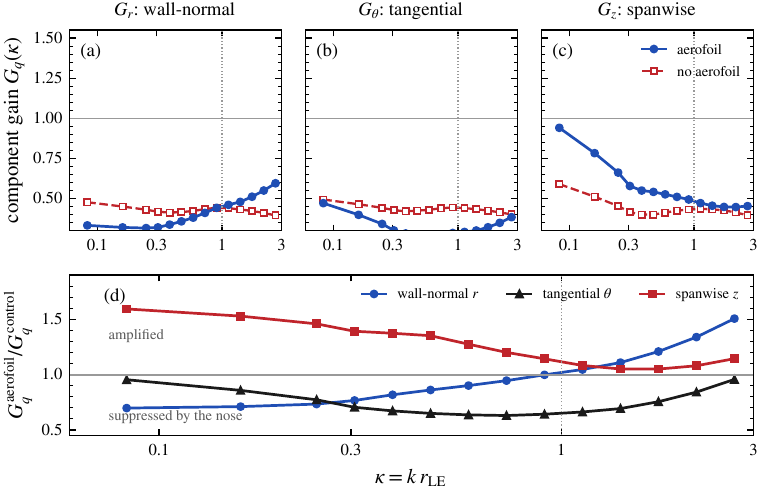}
  \caption{Component gains $G_q(\kappa)$ (\ref{eq:gain}) in the cylindrical
  frame (\ref{eq:projection}): (\textit{a})~wall-normal, (\textit{b})~tangential,
  (\textit{c})~spanwise; aerofoil (solid, filled) versus control (dashed, open).
  (\textit{d})~the same gains referenced to the control, so that free decay
  divides out and departure from unity is the aerofoil's own effect.
  Mesh-independence of the split is documented in the supplementary material. An interactive \href{https://cocalc.ai/share/XRLrBkGfCyJu}{JFM Notebook} is available.}
  \label{fig:tensorial}
\end{figure}

Figure~\ref{fig:tensorial} shows that they do not. At the
energy-containing scales the wall-normal (radial) component is strongly
suppressed,
$G_r\approx0.33$, the tangential moderately, $G_\theta\approx0.47$, while
the spanwise component passes almost untouched, $G_z\approx0.94$. The
no-aerofoil control separates cause from coincidence: without the aerofoil
the three gains are nearly equal ($G_r\approx0.48$, $G_\theta\approx0.49$,
$G_z\approx0.59$ --- ordinary decay, for which a single scalar would nearly
suffice); the aerofoil drives them apart. Towards smaller scales the split
narrows only partially (onset near $\kappa\approx1$, where an eddy first
matches the nose) and remains clearly open across the resolved band.
Panel~(\textit{d}) states the same result in a single quantitative
measure: referenced to the control, so that free decay divides out, the
spanwise gain is amplified by ${\approx}1.6$ at the energy-containing
scales while the wall-normal gain is suppressed to ${\approx}0.7$, and the
components approach one another only near $\kappa=1$. The distortion is
therefore strong, organised by $\kappa$, and irreducibly tensorial.

A gain, however, cannot answer the deeper question. $G_q$ compares energy
levels; it takes the same value whether the pre-impact turbulence is a
deterministic, linearly filtered image of the upstream field or an entirely
different field that merely carries comparable energy. Distinguishing those
two possibilities requires the cross-spectra \emph{between} the stations,
which the congruent surfaces provide.

\subsection{The linearity spectrum, and a low, receding boundary}
\label{sec:results_gamma}
\label{sec:framework_split}

In words, the quantity we now construct asks, scale by scale: \emph{what
fraction of the turbulence arriving at the nose could have been predicted
by the best possible linear formula applied to the turbulence recorded
upstream?} Let
$\mathsf{S}^{ab}_{ij}(\kappa)=\langle\hat u^{a\ast}_i\,\hat u^{b}_j\rangle$
be the cross-spectral tensor between stations $a$ (upstream) and $b$
(downstream), evaluated at matched vertices. The best linear (Wiener)
estimator of the downstream field from the upstream one is
$\mathsf{H}(\kappa)=\mathsf{S}^{ba}(\mathsf{S}^{aa})^{-1}$, and the
fraction of downstream spectral energy it accounts for is the
\emph{input--output linearity spectrum}
\begin{equation}
  \gamma^{2}(\kappa)
  = \frac{\operatorname{tr}\!\big[\mathsf{S}^{ba}\,(\mathsf{S}^{aa})^{-1}\,
          \mathsf{S}^{ab}\big]}
         {\operatorname{tr}\,\mathsf{S}^{bb}}
  \;\in\;[0,1],
  \label{eq:gamma2}
\end{equation}
the multiple coherence of the three velocity components.
$\gamma^{2}\to1$ means a deterministic linear transfer;
$\gamma^{2}\to0$ means the field at that scale is not linearly inherited
from upstream, however similar the two spectral levels may look.
Equivalently, the pre-impact tensor splits exactly into a part the linear
map carries and a residual it cannot,
\begin{equation}
  \mathsf{S}^{bb}
  = \underbrace{\mathsf{H}\,\mathsf{S}^{aa}\,\mathsf{H}^{\dagger}}
    _{\textstyle\text{inherited}}
  \;+\; \underbrace{\mathsf{N}}_{\textstyle\text{generated}},
  \qquad
  \operatorname{tr}\mathsf{N} = \big(1-\gamma^{2}\big)\,\operatorname{tr}\mathsf{S}^{bb},
  \label{eq:decomp}
\end{equation}
a decomposition we will need again in \S\ref{sec:mechanism}, where the
generated tensor $\mathsf{N}$ takes centre stage.

\begin{figure}
  \centering
  \includegraphics[width=0.9\textwidth]{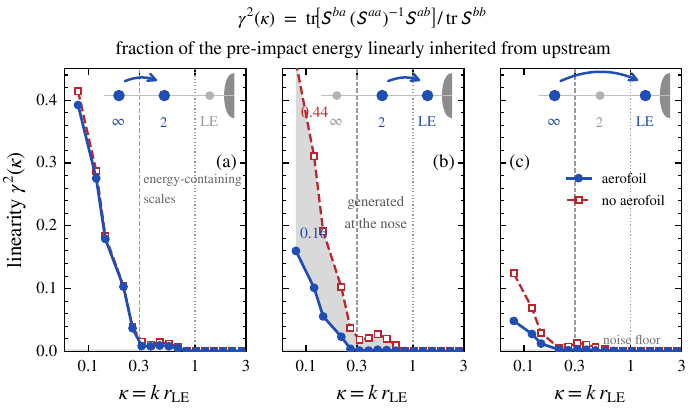}
  \caption{Linearity spectrum $\gamma^{2}(\kappa)$ (\ref{eq:gamma2}) for the
  three station pairs, aerofoil (solid, filled) versus control (dashed, open):
  (\textit{a})~upstream leg, (\textit{b})~near-leading-edge leg,
  (\textit{c})~end to end; the sketch in each panel marks the station pair
  entering that estimate. Grey fill in (\textit{b}): the aerofoil-specific
  loss --- energy generated at the nose rather than inherited. Shaded band
  near zero: empirical null floor; dashed line: $\kappa=0.3$; dotted:
  $\kappa=1$; $U_c\approx18.6~\mathrm{m\,s^{-1}}$. An interactive \href{https://cocalc.ai/share/din5KgyWTpXT}{JFM Notebook} is available.}
  \label{fig:linearity}
\end{figure}

Because everything below leans on this estimator, its statistics are fixed
first. The null level is evaluated from the records, not modelled: repeating the estimate
with the downstream record displaced by a large non-physical lag gives a
floor $\gamma^{2}_{0}\approx3.5\times10^{-3}$, shaded in
figure~\ref{fig:linearity}. The floor is small because the estimate does
not rest on the Welch segments alone --- a three-channel multiple coherence
from $n$ independent averages is biased upward by $\mathcal{O}(3/n)$, which
for ten segments in isolation would be ${\sim}0.3$, but the sector average
spans $0.8$~m of span at a spanwise coherence length of order $14$~mm,
contributing tens of effectively independent samples per segment; the
displaced-record floor is the resulting total bias, quantified. A segment
bootstrap ($400$ draws) puts $95\%$ half-widths of $\pm0.02$--$0.03$ on
$\gamma^{2}$ at the largest scales for every station pair. Features below
the floor are not interpreted.

Figure~\ref{fig:linearity} shows $\gamma^{2}(\kappa)$ for the three station
pairs, and three features organise the result. First, the linear map is
strongest at the largest scales, as rapid-distortion reasoning requires
(small eddies turn over too fast for any linear description to bind them):
on the upstream leg $\gamma^{2}$ rises to $\approx0.39$ at the largest
resolved scale. Second, it is nonetheless bounded well below unity
\emph{everywhere}, and collapses towards the floor already by
$\kappa\approx0.3$ --- at the energy-containing scales, well below the
nose-radius scale $\kappa=1$. Third, the boundary \emph{recedes}
downstream: in figure~\ref{fig:linearity}\textit{b} the blue aerofoil
curve starts at only $\gamma^{2}\approx0.16$ at the largest scales while
the red dashed control starts at $0.44$, and end to end
(figure~\ref{fig:linearity}\textit{c}) the far-upstream field retains
essentially no linear image at the nose ($\gamma^{2}\approx0.05$).

The no-aerofoil control excludes the possibility that this loss is
ordinary convective decorrelation. The station separations are identical
in the two computations, so whatever coherence is lost to convection and
estimator jitter is common to both; yet the near-leading-edge leg retains
$\gamma^{2}\approx0.44$ \emph{without} the aerofoil and only
${\approx}0.16$ with it. The factor-of-three drop below the free-decay
baseline --- with non-overlapping bootstrap intervals --- is the aerofoil's
own doing, localised to the final nose radii of the approach. Resolved by
poloidal sector, the loss is strongest \emph{off} the stagnation line
($\gamma^{2}\approx0.15$ on the shoulders against $0.20$ windward), where
the lateral straining is strongest.
\label{sec:results_boundary}
\label{sec:results_arms}

The statement must be read at the level on which it is computed.
Rapid-distortion corrections act on second-order \emph{spectra}, and a
spectral map may remain serviceable even where $\gamma^{2}$ is small; what
$\gamma^{2}$ bounds is the field-level linear inheritance in
(\ref{eq:decomp}). But the bound explains an otherwise empirical rule of
the prediction literature: turbulence inputs for successful noise models
must be taken close to the leading edge
\citep{deSantana2016,dosSantos2023}, because the far-upstream field cannot
be carried to the nose as a linear datum --- most of what arrives there was
generated on the way. This is the spectral half of the main result. If the loss of linearity is physical rather than an estimator artefact, the same location should appear in ordinary single-point statistics. Section~\ref{sec:results_volume} shows that it does, and identifies which part of the linear description fails.

\subsection{The same boundary in physical space}
\label{sec:results_volume}
\label{sec:framework_descriptions}
\label{sec:framework_rdt}
\label{sec:results_control}

Single-point structure is captured by the Reynolds-stress anisotropy
tensor and its invariants,
\begin{equation}
  b_{ij} = \frac{R_{ij}}{2k_{\mathrm{t}}} - \frac{1}{3}\delta_{ij},
  \qquad
  II = -\tfrac{1}{2}\,b_{ij}b_{ji},
  \qquad
  III = \tfrac{1}{3}\,b_{ij}b_{jk}b_{ki},
  \label{eq:anisotropy}
\end{equation}
with $R_{ij}=\langle u_i'u_j'\rangle$ and
$k_{\mathrm{t}}=\tfrac12 R_{ii}$. Every realisable state lives between two
axisymmetric branches \citep{LumleyNewman1977,Banerjee2007}: $III>0$ means
one dominant eigenvalue (a \emph{prolate}, cigar-like state), $III<0$ one
deficient eigenvalue (an \emph{oblate}, pancake-like state), and the sign
of $III_b\equiv\det\mathsf{b}$ therefore records the \emph{type} of the
anisotropy, independent of its amplitude. Except where noted, every
quantity below is the ratio of the aerofoil case to the control at the same
station, so the natural decay of the grid turbulence divides out and what
remains is the aerofoil's distortion alone.

\begin{figure}
  \centering
  \includegraphics[width=0.9\textwidth]{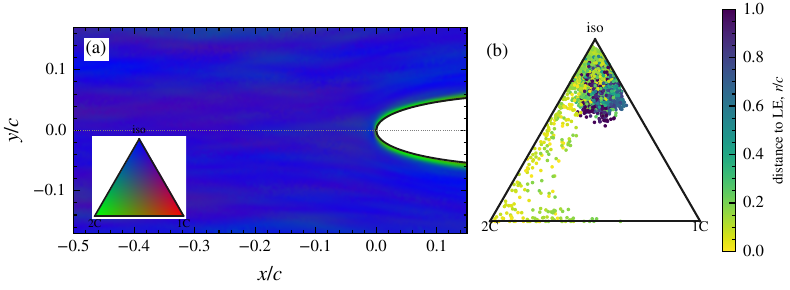}
  \caption{Componentality near the nose (volume midspan plane):
  (\textit{a})~barycentric map (red, one-component; green, two-component; blue,
  isotropic) with the aerofoil outline; (\textit{b})~the same states in the
  barycentric triangle, coloured by distance to the leading edge. An interactive \href{https://cocalc.ai/share/bxQGjyfYH8mF}{JFM Notebook} is available.}
  \label{fig:aniso_map}
\end{figure}

\begin{figure}
  \centering
  \includegraphics[width=\textwidth]{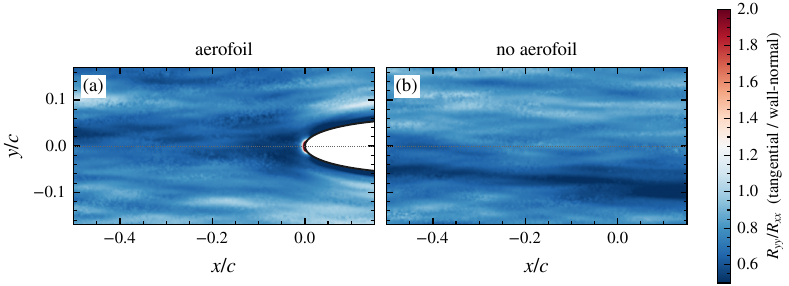}
  \caption{Mean-field footprint of the blocking: $R_{yy}/R_{xx}$ near the
  nose, (\textit{a})~aerofoil, (\textit{b})~control. An interactive \href{https://cocalc.ai/share/QBD8jThdaLfc}{JFM Notebook} is available.}
  \label{fig:blocking_field}
\end{figure}

Figure~\ref{fig:aniso_map} locates the distortion region: in
panel~(\textit{a}) the approach flow renders in the blue of
near-isotropy, and only a compact zone of order $r_{\mathrm{LE}}$ at the
nose shifts towards the green of the two-component (disk) state; the barycentric weights $C_1,C_2,C_3$ used there measure how close
the local state is to one-, two- and three-component turbulence. The mean-field footprint of the blocking is equally direct
(figure~\ref{fig:blocking_field}): on the stagnation streamline the
wall-normal direction is streamwise, so blocking suppresses $R_{xx}$ there,
and the tangential-to-normal ratio $R_{yy}/R_{xx}$ rises to ${\approx}2$ in
a compact region at the nose while the control stays at unity.

\begin{figure}
  \centering
  \includegraphics[width=\textwidth]{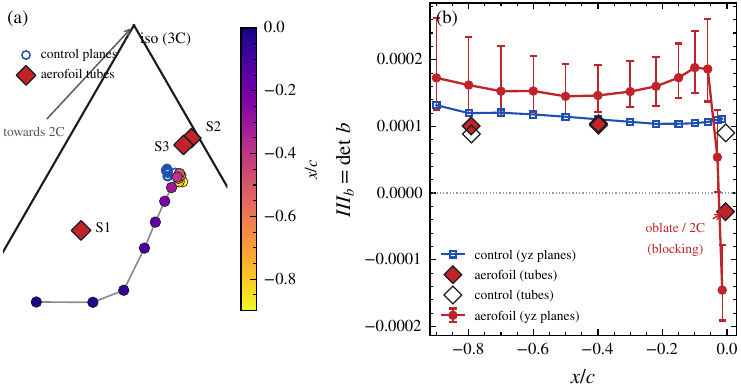}
  \caption{Anisotropy along the stagnation streamline:
  (\textit{a})~barycentric triangle, zoomed to the isotropic apex ($y$--$z$-plane
  tensors coloured by $x/c$; toroidal stations as diamonds);
  (\textit{b})~$III_b=\det b$ against $x/c$, aerofoil (red) versus control
  (blue). An interactive \href{https://cocalc.ai/share/onsFYEE7oksW}{JFM Notebook} is available.}
  \label{fig:aniso_traj}
\end{figure}

Followed along the stagnation streamline (figure~\ref{fig:aniso_traj}),
computed two independent ways that agree --- the mean Reynolds tensor on
the cross-flow planes, and the fluctuation tensor accumulated over the
$24\,769$ snapshots of the windward toroidal strip --- the anisotropy
descends from near-isotropy towards the two-component edge, and the
invariant $III_b$ makes the boundary sharp: prolate from $x/c=-0.9$ to
$-0.06$, it \emph{changes sign} near $x/c\approx-0.026$, while the control
never crosses. The change of type, not merely amplitude, is confined to the
final ${\approx}3\%$ of chord and is caused by the aerofoil.

\begin{figure}
\centering
\resizebox{\textwidth}{!}{%
\begin{tikzpicture}[>=Latex, font=\small]
\begin{scope}[shift={(0,0)}]
  \node at (2.2,2.55) {(\textit{a})\; straining (free space), Eq.~(\ref{eq:worked_mode})};
  \draw[->,gray] (1.85,2.0) -- (1.25,2.0);
  \draw[->,gray] (2.55,2.0) -- (3.15,2.0);
  \draw[->,gray] (2.2,2.30) -- (2.2,2.10);
  \draw[->,gray] (2.2,1.70) -- (2.2,1.90);
  \node[gray] at (2.2,2.0) {$a$};
  \foreach \i/\s in {0/1,1/-1,2/1,3/-1}
    \draw[->,thick] ({0.25+0.38*\i},0.55) -- ++(0,{0.30*\s});
  \draw[->] (0.25,0.10) -- (1.39,0.10);
  \node[below] at (0.82,0.10) {$\boldsymbol{k}_0 = k_0\boldsymbol{e}_t$};
  \node at (0.82,1.18) {$\hat{u}_{0n}$};
  \draw[->,thick] (1.75,0.62) -- (2.45,0.62);
  \node[above] at (2.10,0.65) {$\mathsf{F}(\vartheta)$};
  \foreach \i/\s in {0/1,1/-1,2/1}
    \draw[->,very thick] ({2.75+0.62*\i},0.55) -- ++(0,{0.52*\s});
  \draw[->] (2.75,0.10) -- (4.05,0.10);
  \node[below] at (3.40,0.10) {$\mathrm{e}^{-\vartheta}k_0\,\boldsymbol{e}_t$};
  \node at (3.42,1.35) {$\mathrm{e}^{\vartheta}\hat{u}_{0n}$};
\end{scope}
\begin{scope}[shift={(5.3,0)}]
  \node at (2.1,2.55) {(\textit{b})\; blocking, Eq.~(\ref{eq:blocking})};
  \draw[very thick] (0,0) -- (4.2,0);
  \foreach \xx in {0.2,0.6,...,4.0}
    \draw ({\xx},0) -- ({\xx-0.14},-0.14);
  \draw[thick] (1.1,1.05) circle (0.62);
  \draw[->] (1.55,1.45) arc (40:130:0.62);
  \draw[->,thick] (1.1,0.43) -- (1.1,0.10);
  \node[anchor=west] at (1.22,0.36) {$\boldsymbol{u}^{(d)}\!\cdot\boldsymbol{n}$};
  \draw[dashed] (1.1,-1.05) circle (0.62);
  \node[below] at (1.1,-1.70) {\footnotesize image system, $\nabla\phi^{(s)}$};
  \draw[->,gray,thick] (0.42,0.18) -- (0.02,0.18);
  \draw[->,gray,thick] (1.95,0.18) -- (2.35,0.18);
  \draw[thick] (3.35,0.42) circle (0.20);
  \draw[densely dotted] (2.95,0.20) -- (3.85,0.20);
  \node[above,align=center] at (3.40,0.66)
        {\footnotesize layer $\sim k^{-1}$\\[-3pt]\footnotesize ($\mathrm{e}^{-kn}$)};
\end{scope}
\begin{scope}[shift={(10.9,0)}]
  \node at (2.1,2.55) {(\textit{c})\; expected competition};
  \draw[->] (0,-0.1) -- (0,2.25);
  \node[left] at (0,2.15) {$G_n$};
  \draw[->] (-0.1,0) -- (4.3,0);
  \node[below] at (4.05,-0.02) {$\kappa$ (log)};
  \draw[dashed] (0,1.05) -- (4.2,1.05);
  \node[left] at (0,1.05) {$1$};
  \draw[densely dashed,gray] (0.15,1.72) -- (1.7,1.72);
  \node[gray,above] at (0.92,1.74) {\footnotesize strain only: $\mathrm{e}^{2\vartheta}$};
  \draw[densely dotted] (2.5,0) -- (2.5,1.55);
  \node[below] at (2.5,-0.02) {\footnotesize $\kappa=1$};
  \node[anchor=west,align=left] at (2.58,1.66)
        {\footnotesize $f \approx U_\infty/2\pi r_{\mathrm{LE}}$};
  \draw[very thick]
        (0.15,0.32) -- (1.7,0.36)
        .. controls (2.4,0.45) and (2.8,0.92) .. (4.1,1.02);
  \node[align=center] at (1.0,0.72)
        {\footnotesize blocking-\\[-3pt]\footnotesize dominated};
  \node[align=center] at (3.62,1.30) {\footnotesize $\mathcal{D}\to\mathcal{I}$};
\end{scope}
\end{tikzpicture}%
}
\caption{The two linear mechanisms: (\textit{a})~plane strain amplifies a
  tangential gust's wall-normal amplitude by $\mathrm{e}^{\vartheta}$
  (Eq.~\ref{eq:worked_mode}); (\textit{b})~blocking suppresses the wall-normal
  component of eddies larger than their distance to the surface
  (Eq.~\ref{eq:blocking}); (\textit{c})~their expected competition in the
  wall-normal gain (schematic, not computed).}
\label{fig:framework_mechanisms}
\end{figure}
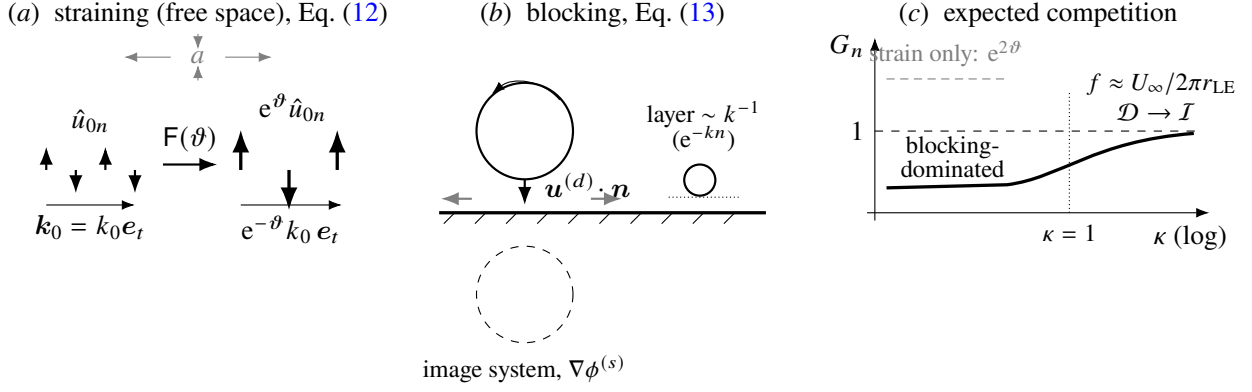

Is this what linear theory predicts? Linear rapid distortion supplies a
concrete benchmark, and we state it rather than re-derive it (full
treatments in \citealp{BatchelorProudman1954,Hunt1973,HuntCarruthers1990}).
Each Fourier mode of weak turbulence evolves independently along mean
trajectories, and the resulting operator acts by congruence,
\begin{equation}
\begin{aligned}
\mathcal{D}(\boldsymbol{k},\vartheta)\big[\Phi^{\infty}\big]
&=
\mathsf{M}(\boldsymbol{k},\vartheta)\,
\Phi^{\infty}\!\big(\mathsf{F}^{\top}\boldsymbol{k},\,\omega\big)\,
\mathsf{M}^{\dagger}(\boldsymbol{k},\vartheta),
\qquad
\mathsf{M}
\sim
\mathsf{P}(\boldsymbol{k})\,\mathsf{F}(\vartheta),
\end{aligned}
\label{eq:rdt_structure}
\end{equation}
with $\mathsf{F}$ the mean deformation gradient and $\mathsf{P}$ the
solenoidal projector. Two consequences matter here. For a transverse gust
aligned with the surface tangent, the classical solution gives
\begin{equation}
  \hat{u}_n = \mathrm{e}^{\vartheta}\,\hat{u}_{0n},
  \qquad
  \hat{u}_z = \hat{u}_{0z},
  \qquad
  \hat{u}_t = 0,
  \label{eq:worked_mode}
\end{equation}
so free-space straining \emph{amplifies} the wall-normal component by
$\mathrm{e}^{2\vartheta}$ in energy. Near a surface, however,
impermeability demands a blocking correction: the velocity field is split
into the strained free-stream part $\boldsymbol{u}^{(d)}$ and a potential
part $\nabla\phi^{(s)}$ chosen so that no fluid crosses the surface,
\citep{Hunt1973,HuntGraham1978},
\begin{equation}
  \frac{\partial\phi^{(s)}}{\partial n}\bigg|_{\mathrm{wall}}
  = -\,\boldsymbol{u}^{(d)}\!\cdot\boldsymbol{n},
  \label{eq:blocking}
\end{equation}
whose influence decays like $\mathrm{e}^{-kn}$: each eddy carries a
blocking layer as thick as its own wavelength, so \emph{large} eddies are
blocked and \emph{small} ones strained, with the crossover set by eddy size
against body radius --- for the aerofoil nose, by
$\kappa=k\,r_{\mathrm{LE}}$, a transposition of Hunt's cylinder result that
recent measurements support directly \citep{dosSantosAIAA2024}.
Figure~\ref{fig:framework_mechanisms} sketches the competition. Because the
linear operator is a congruence it can move the anisotropy \emph{along} the
prolate branch but can never reverse the sign of $III_b$ --- a
constraint that becomes central in \S\ref{sec:mechanism}.
\label{sec:framework_observables}

\begin{figure}
  \centering
  \includegraphics[width=0.9\textwidth]{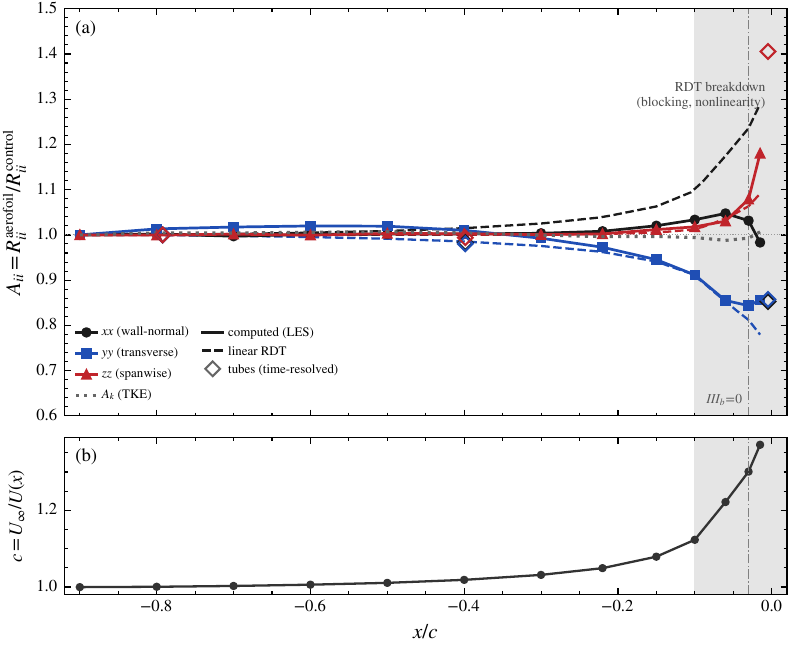}
  \caption{(\textit{a})~Aerofoil/control stress ratios $A_{ii}$ on the
  stagnation streamline (solid; dotted, energy ratio $A_k$; diamonds, toroidal
  stations) against the exact plane-strain rapid-distortion prediction (dashed,
  no fitted parameter); dash-dot line: the $III_b$ crossing.
  (\textit{b})~mean deceleration ratio $U_\infty/U(x)$, which fixes the
  strain. An interactive \href{https://cocalc.ai/share/z9Pfkpjwg6Z3}{JFM Notebook} is available.}
  \label{fig:rdt_boundary}
\end{figure}

Figure~\ref{fig:rdt_boundary} confronts the computation with this
benchmark, with \emph{no fitted parameter}: the exact linear response to
irrotational plane strain, evaluated with the total strain fixed
kinematically by the computed mean deceleration ratio $U_\infty/U(x)$. On the stagnation streamline the leading edge redistributes energy
between components at nearly constant total. The measure of that
redistribution is the stress ratio
\begin{equation}
  A_{ii}(x) =
  \frac{R_{ii}^{\mathrm{aerofoil}}(x)}{R_{ii}^{\mathrm{control}}(x)}
  \qquad\text{(no summation)},
  \label{eq:Aii}
\end{equation}
each normal stress in the aerofoil computation divided by its value in the
no-aerofoil control at the same station, so that the free decay of the
grid turbulence cancels and any departure from unity is the aerofoil's own
effect; $A_k$ denotes the same ratio formed with the kinetic energy. The
total is essentially conserved ($A_k\approx1.02$, the dotted curve of
figure~\ref{fig:rdt_boundary}\textit{a}) while the spanwise stress is
amplified ($A_{zz}\to1.18$, red triangles), the transverse suppressed
($A_{yy}\to0.85$, blue squares) and the wall-normal (black circles)
blocked back towards unity. Against the linear prediction (the dashed curves of
figure~\ref{fig:rdt_boundary}\textit{a}), the transverse component
(blue squares) follows RDT essentially to the nose; the wall-normal
component (black circles) follows it to $x/c\approx-0.2$ and then falls
far below its steeply rising dashed counterpart (linear strain would
amplify it --- the surface blocks it instead); the spanwise component
rises above prediction only in the final few per cent of chord. The linear
description is quantitatively accurate for $x/c\lesssim-0.1$ and fails
inside that band, component by component, at the same location as the
$III_b$ sign change.

Two features of this comparison repay scrutiny, and both, once examined,
reinforce it. First, the benchmark is free-space strain: it omits blocking, so the
wall-normal shortfall \emph{is} the blocking, quantified. Second, RDT is a
linearisation, and the computed strain field (supplementary material)
shows it extrapolated to $\vartheta=\ln(U_\infty/U)\approx0.68$ at the
innermost station, with the rapid-distortion parameter
$S_r=a\,\tau_{\mathrm{turnover}}$ (strain rate times eddy-turnover time)
exceeding unity only over the final few
per cent of chord --- precisely the breakdown band. The distortion there is
simultaneously strong and \emph{rapid}, so its failure is not slow
nonlinear relaxation, which rapidity excludes.

The spectral and physical-space observations are consistent: linear inheritance collapses at the energy-containing scales, and the anisotropy changes type over the final approach, departing a parameter-free linear benchmark.
A quantitative implication of (\ref{eq:worked_mode}) is that free-space straining would amplify the wall-normal component by $\mathrm{e}^{2\vartheta}$; the observed suppression therefore means that blocking has more than offset that amplification.

\subsection{The distortion at second order: a well-defined operator}
\label{sec:results_operator}

Low field-level linearity does not make the distortion useless to a noise
model. A noise model consumes only second-order quantities --- the upwash
spectrum and its spanwise scale --- and at that level the distortion is a
well-defined spectral map even where the phase-resolved inheritance is
small: $\gamma^{2}$ bounds the realisation, the operator maps the spectra.
We therefore deliver the acoustically relevant piece in usable form.

\begin{figure}
  \centering
  \includegraphics[width=0.66\textwidth]{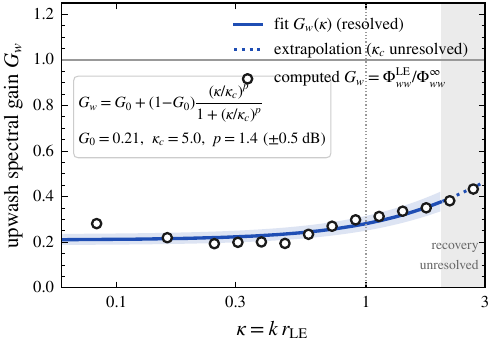}
  \caption{Upwash gain
  $G_v(\kappa)=\Phi_{vv}^{\mathrm{LE}}/\Phi_{vv}^{\infty}$ (symbols) with the
  empirical fit (\ref{eq:blockfit}) (line; shading, $\pm0.5$~dB r.m.s. fit
  error). The recovery scale $\kappa_c$ lies beyond the resolved band. An interactive \href{https://cocalc.ai/share/8712wKgPMbMK}{JFM Notebook} is available.}
  \label{fig:operator}
\end{figure}

Figure~\ref{fig:operator} reports the upwash gain $G_v(\kappa)$ --- the gain of the transverse
(upwash) component, the one that loads the aerofoil --- over the
resolved band: a flat blocking, $G_v\approx0.21$ ($-6.8$~dB), at the
energy-containing scales, recovering partially from $\kappa\approx1$ to
$G_v\approx0.44$ by the resolution limit. The three-parameter fit
\begin{equation}
  G_v(\kappa) \;=\; G_0 + (1-G_0)\,
  \frac{(\kappa/\kappa_c)^{p}}{1+(\kappa/\kappa_c)^{p}},
  \qquad
  G_0 = 0.21,\quad \kappa_c = 5.0,\quad p = 1.43,
  \label{eq:blockfit}
\end{equation}
reproduces the computed gain to $0.5$~dB r.m.s. It is an empirical
interpolation whose only theoretical content is its two limits (complete
blocking as $\kappa\to0$, free passage as $\kappa\to\infty$); the blocking
level and the recovery \emph{onset} are firmly resolved, the recovery's
completion lies beyond the resolved range ($\kappa_c$ interpolates, it is
not an independently determined scale). Its domain of validity is inherited from
\S\ref{sec:results_gamma}: this is a spectral map, valid for the
second-order inputs of a noise model, explicitly not a field-level
propagator --- through (\ref{eq:decomp}) the pre-impact energy is
${\gtrsim}95\%$ generated in situ end to end.

Most of the turbulence that reaches the leading edge is therefore generated locally rather than transported from upstream. The generation changes the sign of $III_b$, something a linear congruence cannot do. The Reynolds-stress budget contains one term capable of the inversion; the next section evaluates it.

\section{The generating mechanism: pressure--strain redistribution}
\label{sec:mechanism}
\label{sec:framework_budget}
\label{sec:framework_ps}

Section~\ref{sec:results} established \emph{that} most of the pre-impact
tensor is generated rather than inherited, and that the anisotropy changes type on the way in, but it did not
identify the dynamics responsible. This section names the
responsible dynamics and establishes it three independent ways: evaluated
directly on the stagnation line, switched on exactly at the anisotropy
inversion, and imprinted on the generated spectral tensor. A four-aerofoil
sweep then tests whether any of it is particular to one geometry.

\subsection{Why only pressure--strain can invert the anisotropy}

\begin{figure}
  \centering
  \resizebox{0.98\textwidth}{!}{%
  \begin{tikzpicture}[>=Latex, font=\small]
    \def\bw{0.26}
    \node[font=\bfseries] at (1.8,4.35) {incident};
    \draw[fill=black!8] (1.8,3.15) circle (0.5); \node[gray] at (1.8,3.15) {iso};
    \fill[cn] (1.41,0.6) rectangle ++(\bw,0.9);
    \fill[ct] (1.77,0.6) rectangle ++(\bw,0.9);
    \fill[cz] (2.13,0.6) rectangle ++(\bw,0.9);
    \node[cn] at (1.54,0.36) {$n$}; \node[ct] at (1.90,0.36) {$t$}; \node[cz] at (2.26,0.36) {$z$};
    \node at (1.8,0.0) {$III_b\approx0$};
    \node[font=\bfseries] at (6.2,4.35) {strained};
    \draw[fill=cn!10,draw=cn,thick] (6.2,3.15) ellipse (0.72 and 0.30);
    \fill[cn] (5.81,0.6) rectangle ++(\bw,1.5);
    \fill[ct] (6.17,0.6) rectangle ++(\bw,0.72);
    \fill[cz] (6.53,0.6) rectangle ++(\bw,0.82);
    \node[cn] at (5.94,0.36) {$n$}; \node[ct] at (6.30,0.36) {$t$}; \node[cz] at (6.66,0.36) {$z$};
    \node[cn] at (6.2,0.0) {prolate, $III_b>0$};
    \node[font=\bfseries] at (10.6,4.35) {pre-impact};
    \draw[fill=cz!10,draw=cz,thick] (10.6,3.15) ellipse (0.34 and 0.62);
    \fill[cn] (10.21,0.6) rectangle ++(\bw,0.82);
    \fill[ct] (10.57,0.6) rectangle ++(\bw,1.15);
    \fill[cz] (10.93,0.6) rectangle ++(\bw,1.30);
    \node[cn] at (10.34,0.36) {$n$}; \node[ct] at (10.70,0.36) {$t$}; \node[cz] at (11.06,0.36) {$z$};
    \node[cz] at (10.6,0.0) {oblate, $III_b<0$};
    \draw[->,very thick,cn] (2.55,3.15) -- (5.35,3.15);
    \node[cn] at (3.95,3.92) {mean strain};
    \node[cn] at (3.95,3.55) {production $P_{nn}>0$};
    \node[cn,font=\itshape] at (3.95,2.55) {amplify $n$};
    \draw[->,very thick,cz] (7.05,3.15) -- (9.9,3.15);
    \node[cz] at (8.48,3.92) {pressure--strain $\Pi$};
    \node[cz] at (8.48,3.55) {return to isotropy};
    \draw[->,ct,thick] (8.48,2.85) -- (8.05,2.35);
    \draw[->,cz,thick] (8.48,2.85) -- (8.91,2.35);
    \node[font=\itshape] at (8.48,2.15) {$n\!\to\!(t,z)$};
  \end{tikzpicture}}
  \caption{The budget mechanism (schematic): production amplifies the
  wall-normal ($n$) stress, driving the state prolate; pressure--strain drains
  it into the tangential ($t$) and spanwise ($z$) components, carrying the state
  oblate. Bars: the three normal stresses.}
  \label{fig:budget}
\end{figure}
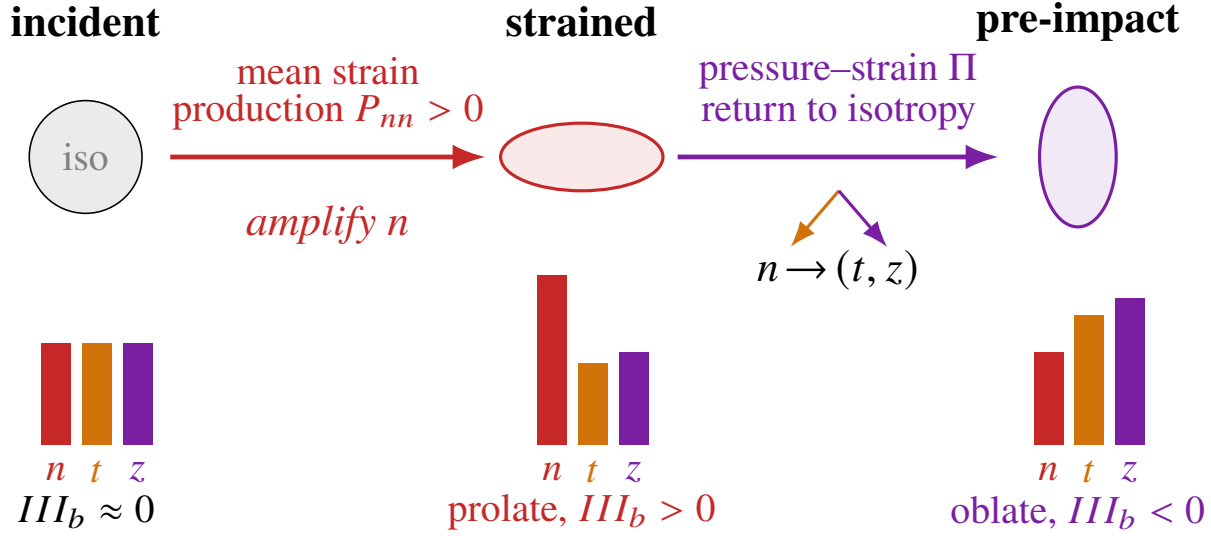

On the windward stagnation line, symmetry makes the wall-normal direction
streamwise; we attach the local frame $n\equiv x$, $t\equiv y$, $z$
spanwise. The transport of the diagonal Reynolds stresses in a stationary
flow reads \citep[e.g.][]{Pope2000}
\begin{equation}
  \underbrace{U_k\,\partial_k R_{ii}}_{\text{convection }C_{ii}}
  \;=\;
  \underbrace{-2\,R_{ik}\,\partial_k U_i}_{\text{production }P_{ii}}
  \;+\;
  \underbrace{\frac{2}{\rho}\big\langle p'\,\partial_i u_i'\big\rangle}_{\text{pressure--strain }\Pi_{ii}}
  \;+\;
  T_{ii} \;-\; \varepsilon_{ii},
  \label{eq:budget}
\end{equation}
(no summation on $i$), with $\rho$ the density, $T_{ii}$ the transport
terms and $\varepsilon_{ii}$ the dissipation. On the stagnation line the mean strain
is plane --- streamwise deceleration $S\equiv\partial_x U_x<0$ balanced by
lateral divergence --- so the production is diagonal,
\begin{equation}
  P_{nn}=-2R_{nn}\,S>0,\qquad
  P_{tt}=+2R_{tt}\,S<0,\qquad
  P_{zz}\approx0,
  \label{eq:production_budget}
\end{equation}
amplifying the wall-normal stress and depleting the tangential one ---
driving the state \emph{prolate}, exactly as linear theory says
(figure~\ref{fig:budget}). Production, however, is bound by the constraint established in
\S\ref{sec:results_volume}: a congruence preserves the sign of
$III_b$.
Production is the single-point image of the linear straining; it cannot
flip the state to oblate. Of the remaining terms in (\ref{eq:budget}),
transport moves energy in space without changing its componentality and
dissipation is close to isotropic in the rapid stagnation zone. 
The only
term built to move energy \emph{between} components at constant total ---
and is therefore the only term that can take the state from prolate to oblate. Pressure--strain is the return-to-isotropy term of single-point closures \citep{SpezialeSarkarGatski1991,DurbinReif2017} and has previously been identified as balancing the mean strain in stagnation-point budgets \citep{XiongLele2007}. We evaluate it directly.

\subsection{Evaluating pressure--strain on a single surface}
\label{sec:method_meas}

Pressure--strain is among the most difficult turbulence quantities to
obtain: it requires the fluctuating pressure and the fluctuating velocity
\emph{gradients}, time-resolved, throughout a volume --- which is why it
is usually modelled rather than evaluated. A traceless identity reduces
the requirement to a single surface. For an
incompressible fluctuation field the tensor is exactly traceless,
\begin{equation}
  \Pi_{ii}=\frac{2}{\rho}\big\langle p'\,\partial_i u_i'\big\rangle
  =\frac{2}{\rho}\big\langle p'\,\nabla\!\cdot\!\boldsymbol{u}'\big\rangle=0.
  \label{eq:traceless}
\end{equation}
The two in-surface components need only tangential and spanwise
derivatives, both available on a sampling surface enclosing the nose,
\begin{equation}
  \Pi_{tt}=\frac{2}{\rho}\big\langle p'\,\partial_t u_t'\big\rangle,\qquad
  \Pi_{zz}=\frac{2}{\rho}\big\langle p'\,\partial_z u_z'\big\rangle,
  \label{eq:pi_intwo}
\end{equation}
and tracelessness then supplies the one component that would otherwise
require an off-surface gradient:
\begin{equation}
  \Pi_{nn}=-\big(\Pi_{tt}+\Pi_{zz}\big).
  \label{eq:pi_nn}
\end{equation}
Concretely, on each toroidal surface $\Pi_{zz}$ is formed from a spanwise
central difference and $\Pi_{tt}$ from a three-point poloidal stencil
along the windward line, averaged over the span and the full
$24\,769$-snapshot record. Two convergence checks fix the robustness of
the estimate: temporally, the spanwise-averaged values are insensitive to
a twelvefold subsampling of the record --- identical to four significant
digits --- and spatially, doubling the width of both derivative stencils
changes $\Pi_{tt}$ by $+4.2\%$ and $\Pi_{zz}$ by less than $0.1\%$.
Where the surfaces are too widely spaced to resolve the near-nose
gradients, an independent second estimate comes from the mean-field
budget on the cross-flow ladder: the deviatoric residual
$\Pi_{ii}\approx\mathrm{dev}(C_{ii}-P_{ii})$, read for its sign structure
and switch-on location only, since the budget does not close exactly under
wall-modelled LES.

\subsection{The redistribution on the stagnation line}
\label{sec:results_pi}

\begin{figure}
  \centering
  \includegraphics[width=0.9\textwidth]{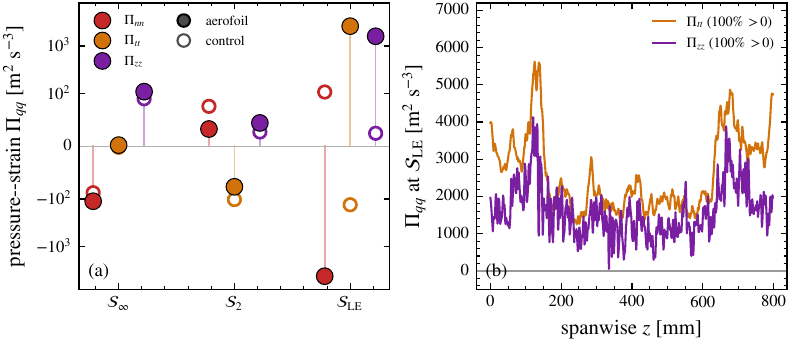}
  \caption{Direct pressure--strain evaluation: (\textit{a})~the three
  diagonal components at the three stations, aerofoil (filled) versus control
  (open); (\textit{b})~spanwise profiles of $\Pi_{tt}$ and $\Pi_{zz}$ at
  $\mathcal{S}_{\mathrm{LE}}$. An interactive \href{https://cocalc.ai/share/Eke6Q3rNt29u}{JFM Notebook} is available.}
  \label{fig:direct_pi}
\end{figure}

At the pre-impact station the diagonal pressure--strain has a strong,
definite sign structure: the wall-normal component is a large
\emph{source} being drained, the tangential and spanwise components are the
\emph{sinks} being filled,
\begin{equation}
  \Pi_{nn}\approx-4.2\times10^{3},\qquad
  \Pi_{tt}\approx+2.6\times10^{3},\qquad
  \Pi_{zz}\approx+1.6\times10^{3}\ \ \mathrm{m^{2}\,s^{-3}},
  \label{eq:pi_measured}
\end{equation}
with the two directly evaluated terms positive at every one of the $799$
resolved spanwise stations (figure~\ref{fig:direct_pi}\textit{b}: the
orange $\Pi_{tt}$ and purple $\Pi_{zz}$ traces remain above zero along
the entire span) --- the redistribution is spanwise-coherent, not a local
event. It is also the aerofoil's own: in panel~(\textit{a}) the filled
symbols at $\mathcal{S}_{\mathrm{LE}}$ stand a factor of $40$--$65$
above the open control symbols, while at the two upstream stations
filled and open symbols coincide at background level. The pattern is exactly a return towards isotropy driven through the
mean strain: production pumps the wall-normal stress up
(\ref{eq:production_budget}); pressure--strain drains it
(\ref{eq:pi_nn}) into the tangential and spanwise components
(figure~\ref{fig:budget}). That the \emph{spanwise} component --- untouched
by the two-dimensional production and the least blocked of the three --- is one of the two sinks is a first indication of a result that recurs
below: whatever turbulence this mechanism generates will be
spanwise-rich. The construction
is explicit about its one inference: the wall-normal term follows from
tracelessness rather than independent evaluation, resting on
incompressibility (well satisfied at $M=0.058$) and on the two directly
evaluated, span-uniform in-surface terms.

\subsection{The redistribution switches on at the anisotropy inversion}
\label{sec:results_switch}

\begin{figure}
  \centering
  \includegraphics[width=0.9\textwidth]{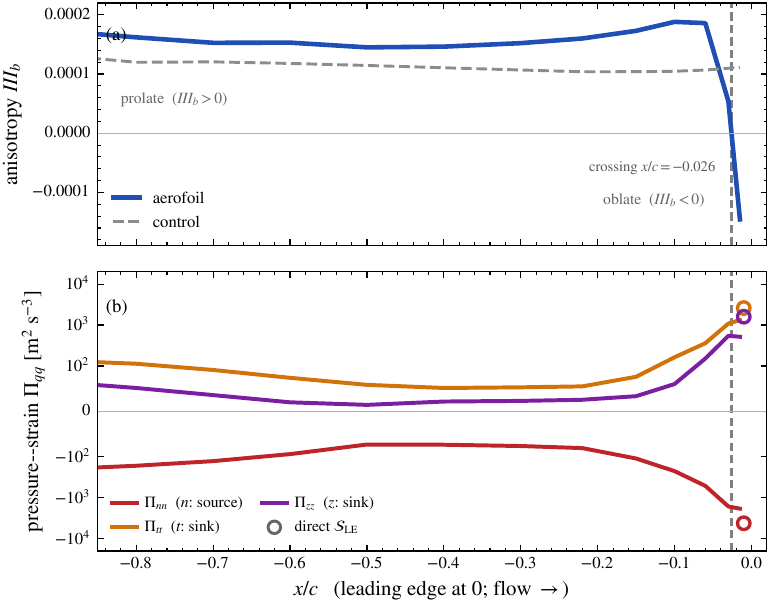}
  \caption{Redistribution and inversion on a shared $x/c$ axis:
  (\textit{a})~$III_b(x/c)$, aerofoil (blue, crossing at $x/c\approx-0.026$)
  versus control (dashed, no crossing); (\textit{b})~diagonal pressure--strain
  (symlog): budget residual (curves) and direct surface value at
  $\mathcal{S}_{\mathrm{LE}}$ (open markers). Dashed vertical line: the
  crossing. An interactive \href{https://cocalc.ai/share/Woo7bFWsmSNq}{JFM Notebook} is available.}
  \label{fig:pi_iiib}
\end{figure}

Figure~\ref{fig:pi_iiib} places the redistribution and the anisotropy on a common $x/c$ axis, showing that the two observations coincide. $III_b$ holds prolate from far upstream
and crosses zero at $x/c\approx-0.026$; the control never crosses. The
pressure--strain --- both the direct surface estimate and the independent
budget residual --- lies at background level while $III_b>0$ and switches
on steeply over the final ${\sim}10\%$ of chord. The two estimates agree
in sign and in switch-on location, and the ordering is causal rather than
merely coincident: the redistribution rises from the background well
upstream of the crossing at $-0.026$ --- the drain precedes the inversion
it produces. (They differ in magnitude near the nose by a factor of two to
three; the band-averaged budget smooths the sharp peak the surface probe
captures one nose radius from the tip.) Two remarks fix the
interpretation. The switch-on is not at the geometric scale $\kappa=1$ but
at the physical location where the accumulated strain becomes rapid
($S_r>1$ over the same final few per cent of chord): this is a rapid,
strong-strain effect, not slow nonlinear relaxation. And both crossing and
switch-on vanish without the aerofoil: neither is a property the incident
turbulence carries in.

\subsection{The generated spectral tensor}
\label{sec:results_N}

\begin{figure}
  \centering
  \includegraphics[width=0.9\textwidth]{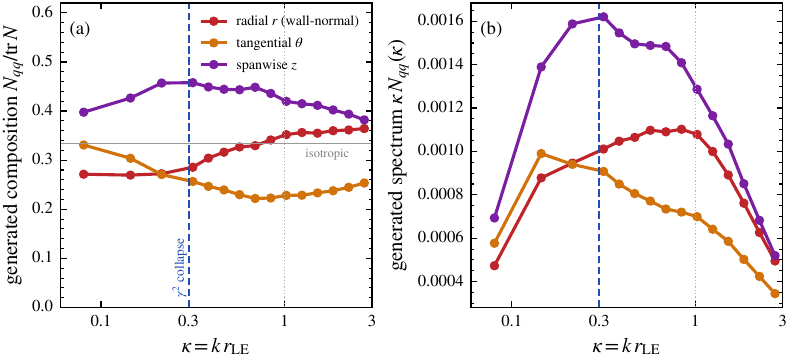}
  \caption{The generated tensor
  $\mathsf{N}=\Phi^{\mathrm{LE}}-\mathsf{H}\Phi^{\infty}\mathsf{H}^{\dagger}$,
  end to end: (\textit{a})~composition $N_{qq}/\operatorname{tr}\mathsf{N}$;
  (\textit{b})~premultiplied spectra $\kappa N_{qq}(\kappa)$. An interactive \href{https://cocalc.ai/share/Woo7bFWsmSNq}{JFM Notebook} is available.}
  \label{fig:generated}
\end{figure}

If pressure--strain generates the pre-impact turbulence, its signature
must appear in the generated tensor $\mathsf{N}$ of
(\ref{eq:decomp}) --- large-scale (the linear inheritance dies first at
the energy-containing scales) and spanwise-rich (the spanwise sink).
Figure~\ref{fig:generated} confirms both: the purple spanwise curve of
panel~(\textit{a}) rises to
$N_{zz}/\operatorname{tr}\mathsf{N}\approx0.46$ against an isotropic
$1/3$ (grey line), peaking at $\kappa\approx0.3$ --- the very scale at
which $\gamma^{2}$ collapses --- while the red and orange in-plane
fractions fall correspondingly below their isotropic share. Repeating the decomposition on the
near-leading-edge leg alone, which isolates the final
$25\,r_{\mathrm{LE}}$ where the aerofoil acts, yields the same
spanwise-dominated composition with negligible inheritance: the character
of the generated turbulence is set by the aerofoil-local redistribution,
independent of how far upstream the reference is taken.

\begin{figure}
  \centering
  \includegraphics[width=0.9\textwidth]{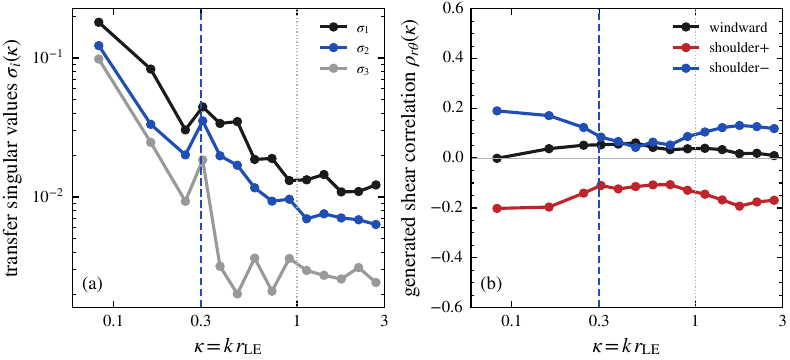}
  \caption{(\textit{a})~Singular values of the transfer
  $\mathsf{H}(\kappa)$. (\textit{b})~generated shear correlation
  $\rho_{r\theta}(\kappa)$ by poloidal sector: opposite signs on the two
  shoulders, near zero on the stagnation line. An interactive \href{https://cocalc.ai/share/XSzrmoDZigop}{JFM Notebook} is available.}
  \label{fig:transfer_full}
\end{figure}

The same cross-spectra resolve the transfer and the generation in full
(figure~\ref{fig:transfer_full}). The singular values of the Wiener
transfer $\mathsf{H}(\kappa)$ are all small ($\sigma_1\approx0.2$ at the
energy-containing scales, below $0.02$ by $\kappa=1$) and of comparable
magnitude --- the weak linear inheritance is spread across the three
velocity modes, not carried by one channel --- and each decays towards the
small scales, as rapid distortion requires. The generation is also tensorial. Resolved by poloidal
sector, the generated shear correlation
\begin{equation}
  \rho_{r\theta}(\kappa)
  = \frac{\operatorname{Re} N_{r\theta}}
          {\left(N_{rr}\,N_{\theta\theta}\right)^{1/2}},
  \label{eq:rho_rtheta}
\end{equation}
the normalised correlation between the wall-normal and tangential
components of the generated field, is negligible on the windward line, as
its up--down symmetry demands. On the two shoulders it reaches
${\approx}\mp0.15$ with \emph{opposite} sign
(figure~\ref{fig:transfer_full}\textit{b}: the red and blue shoulder
curves are mirror images about zero while the black windward curve stays
at the axis) --- an antisymmetric shear stress that cancels only under
the azimuthal average. The aerofoil generates the full tensor, off-diagonal
correlations included.

\subsection{Nose-radius scaling and a bluntness threshold}
\label{sec:results_sweep}

The results so far derive from a single leading-edge geometry. We
therefore repeat the analysis on a four-aerofoil family spanning a fourteen-fold range of
nose radius --- NACA~66-006, 0008, 0012, 0016;
$r_{\mathrm{LE}}=0.8$--$11.2$~mm --- each an otherwise identical
${\sim}1.1\times10^{9}$-node computation (table~\ref{tab:matrix}).

\begin{figure}
  \centering
  \includegraphics[width=0.9\textwidth]{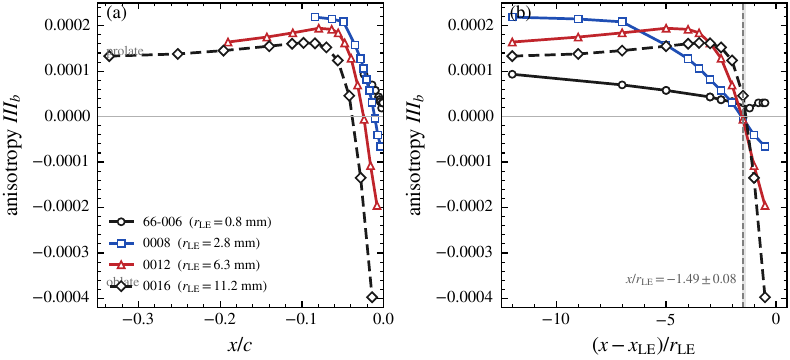}
  \caption{The crossing across the sweep: $III_b$ on the windward approach
  (\textit{a})~against $x/c$ and (\textit{b})~against
  $(x-x_{\mathrm{LE}})/r_{\mathrm{LE}}$, where the blunt sections collapse at
  $-1.49\pm0.08$ (grey band) and the sharp 66-006 stays prolate. Plane-averaged
  over $|y-y_{\mathrm{LE}}|<20$~mm. An interactive \href{https://cocalc.ai/share/nd2Uvxct8a4r}{JFM Notebook} is available.}
  \label{fig:sweep_iiib}
\end{figure}

The anisotropy crossing is the sharpest test, and it collapses on the nose
radius (figure~\ref{fig:sweep_iiib}). In chord units (figure~\ref{fig:sweep_iiib}\textit{a}) the crossing
marches upstream as the nose blunts, from $x/c\approx-0.011$ for the
0008 (blue squares) to $-0.038$ for the 0016 (black diamonds); rescaled
by $r_{\mathrm{LE}}$ in panel~(\textit{b}) the three blunt aerofoils
coincide at $x/r_{\mathrm{LE}}=-1.49\pm0.08$, the 0012 value
agreeing with the finer-ladder estimate of figure~\ref{fig:pi_iiib} to
within the station spacing. It is the nose region, not the chord, that
sets where the anisotropy inverts. One distinction must be drawn with
care: within the four-digit family $r_{\mathrm{LE}}=1.1019\,t_c^{2}c$, so this collapse
cannot by itself distinguish nose radius from any monotone function of
thickness --- it excludes the chord, and separating the two requires a
section from outside the family. The sharp-nosed 66-006 is that outside
case, and it is the informative exception: its trajectory dips to
$III_b\approx+0.2\times10^{-4}$ at $x/r_{\mathrm{LE}}\approx-1.2$ --- a
fifth of its upstream level, essentially zero on the scale of the figure
--- and recovers without reaching the oblate side: a bluntness threshold
below which the accumulated windward strain is too weak to complete the
inversion, an interpretation we hold provisionally.

\begin{figure}
  \centering
  \includegraphics[width=0.9\textwidth]{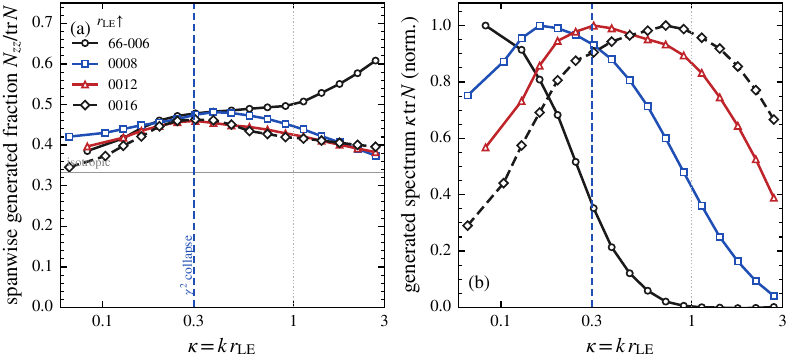}
  \caption{Generated tensor across the sweep (near-leading-edge leg):
  (\textit{a})~spanwise fraction $N_{zz}/\operatorname{tr}\mathsf{N}$;
  (\textit{b})~premultiplied $\kappa\operatorname{tr}\mathsf{N}$, each
  normalised to its peak. An interactive \href{https://cocalc.ai/share/xbNGmGoyWy5r}{JFM Notebook} is available.}
  \label{fig:sweep_generated}
\end{figure}

The generated composition, by contrast, is the sweep's cleanest
invariant (figure~\ref{fig:sweep_generated}): the spanwise share of the
generated energy is $0.45,\,0.46,\,0.44,\,0.43$ across the four sections
--- a few per cent spread about $0.45$, against an isotropic $1/3$ ---
even for the 66-006 whose anisotropy never fully inverts. The linearity
level itself does not collapse on $\kappa$ (figure in the supplementary
material): it falls with $r_{\mathrm{LE}}$ because the stations sit at
fixed multiples of the nose radius, so larger noses mean longer transits
and more decorrelation --- a separation effect, not a nose effect; only
the relative near-edge collapse is common to the family. The sweep's
conclusion is thus precise: \emph{where} the mechanism acts scales
with the
nose radius, and \emph{what} it builds --- spanwise-rich, large-scale
turbulence --- is universal across the family: whenever the leading edge
generates turbulence, it generates it spanwise-rich.

The redistribution also provides single-point closure with a concrete
test: the SSG model \citep{SpezialeSarkarGatski1991}, evaluated from the same mean
fields, reproduces the computed redistribution in sign everywhere and in
magnitude to within ${\sim}15\%$ over the outer approach, but
under-predicts the sharp, strain-gated near-nose intensification by a
factor of about two (figure in the supplementary material) --- a challenge
these data pose to rapid pressure--strain closures.
\label{sec:results_ssg}

\subsection{One mechanism, both boundaries}
\label{sec:results_synth}

The spectral and physical-space observations share a common cause.
Pressure--strain redistribution generates turbulence locally --- hence the
collapse of $\gamma^{2}$, since a field created in the distortion region
cannot be a linear image of the upstream one --- and it does so by
draining the strain-amplified wall-normal stress into the tangential and
spanwise components --- hence the $III_b$ crossing, prolate to oblate. The
spectral collapse and the physical-space inversion are two projections of
one process, and the turbulence that process leaves at the edge is
spanwise-rich and large-scale. The spanwise-rich character is
acoustically decisive: the spanwise coherence of the upwash is one of the
two quantities that set radiated leading-edge noise, so the mechanism
identified here has a direct, quantifiable acoustic consequence. The final section evaluates it against
experiment.

\section{The acoustic consequence}
\label{sec:results_acoustic}

An Amiet-type prediction factorises the far-field spectral density as
\begin{equation}
  S_{pp}(\omega) \;\propto\;
  \ell_z(\omega)\,\Phi_{vv}(\omega)\,|\mathcal{L}(\omega)|^{2},
  \label{eq:amiet}
\end{equation}
the product of the spanwise coherence length of the upwash, its spectrum,
and the flat-plate aeroacoustic transfer function \citep{Amiet1975}:
physically, how long a strip of the edge is forced coherently, how
strongly it is forced, and how efficiently that forcing radiates. The
near-nose distortion enters only through the first two factors, both of
which the aligned surfaces quantify directly, so the ratio of the
distorted to the undistorted prediction,
\begin{equation}
  \mathcal{R}(\omega) = G_v(\omega)\,\frac{\ell_z^{\mathrm{LE}}(\omega)}
  {\ell_z^{\infty}(\omega)},
  \label{eq:acoustic_ratio}
\end{equation}
is independent of the transfer function --- a statement of what the
distortion does to the radiated field that involves no acoustic modelling
beyond the factorisation itself. Two computed effects enter it with
opposite sign.

\subsection{Two computed effects of opposite sign}
\label{sec:results_ratio}

\begin{figure}
  \centering
  \includegraphics[width=\textwidth]{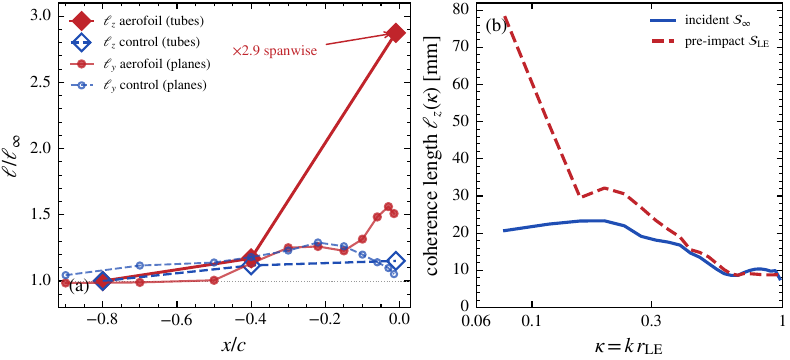}
  \caption{Growth of the incident length scales:
  (\textit{a})~$\ell_z$ (toroidal stations) and $\ell_y$ ($y$--$z$ planes)
  against $x/c$, aerofoil versus control; (\textit{b})~frequency-resolved
  spanwise coherence length $\ell_z(\kappa)$ on the stagnation line at the three
  stations. An interactive \href{https://cocalc.ai/share/QNPpdkSVdmwL}{JFM Notebook} is available.}
  \label{fig:scales}
\end{figure}

The first is the upwash blocking already delivered as
(\ref{eq:blockfit}): $-6.8$~dB across the energy-containing scales. The
second is structural, and it is the acoustic face of the mechanism of
\S\ref{sec:mechanism}. The spanwise coherence length of the upwash is
\emph{stretched} by the aerofoil (figure~\ref{fig:scales}). Its
\emph{integral} value grows from $\ell_z\approx14$~mm at the incident
station to ${\approx}40$~mm at the pre-impact station --- a factor
${\approx}2.9$ (panel~\textit{a}) --- while the control stays flat.
Resolved by scale (panel~\textit{b}), the stretching is concentrated in
the energetic, low-$\kappa$ eddies and leaves the small scales
($\kappa\to1$) untouched. A lateral-scale increase of $1.4$--$1.8\times$ close to the
nose has been measured by hot wire \citep{dosSantos2023}; the
frequency-resolved, component-specific stretching quantified here is its
spectral form. The blocking flattens the approaching eddies against the nose,
and the spanwise-rich generated turbulence carries wider coherence.
Because radiated power scales linearly with $\ell_z$, this stretching \emph{raises} the predicted noise and so opposes the
blocking: at the very
largest scales $+5.5$~dB of stretching nearly cancels $-6.8$~dB of
blocking (net ${\approx}-1$~dB), and the cancellation weakens with scale
until the net correction approaches the full blocking by
$\kappa\approx0.3$--$0.5$. No scalar correction --- carrying $G_v$ but not
$\ell_z$ --- can represent this structure.

\subsection{Prediction of the laboratory experiment}

\begin{figure}
  \centering
  \includegraphics[width=0.9\textwidth]{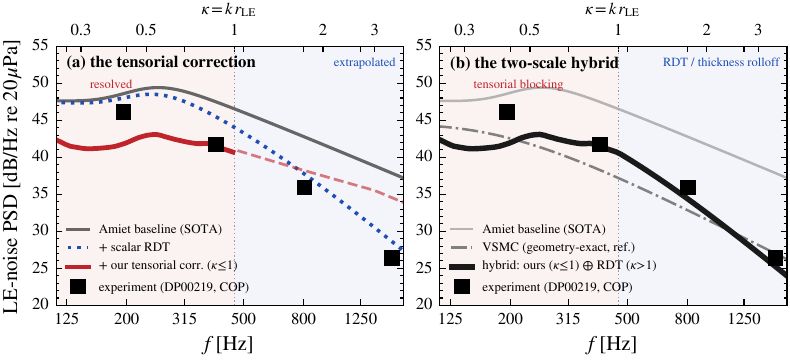}
  \caption{Prediction against the laboratory measurement (markers):
  (\textit{a})~Amiet baseline, scalar rapid-distortion filter, and the computed
  tensorial correction $\mathcal{R}(\kappa)$ (solid; extrapolated dashed);
  (\textit{b})~two-scale hybrid (tensorial for $\kappa\le1$, scalar roll-off for
  $\kappa>1$) with an independent vortex-particle computation
  \citep{SharmaSarradjSchmidt2020} for reference. Shading: the two scale
  regimes split at $\kappa=1$. An interactive \href{https://cocalc.ai/share/ZEqbHxcQrkA2}{JFM Notebook} is available.}
  \label{fig:acoustic_hybrid}
\end{figure}

The comparison with experiment uses the laboratory configuration of
\S\ref{sec:ExpSetup}: the measured, isolated leading-edge noise of the
NACA~0012 in grid turbulence.
\label{sec:results_prediction}
Applied as a level correction to a standard Amiet prediction
\citep{Amiet1975,PatersonAmiet1976}, the computed $\mathcal{R}(\kappa)$ (red curve of
figure~\ref{fig:acoustic_hybrid}\textit{a}) brings the grey flat-plate
baseline --- which runs a mean ${+}6.6$~dB above the measured symbols ---
onto the measured spectrum over the energy-containing band
$\kappa\lesssim1$, to a mean absolute deviation of $2.3$~dB, from
simulation-derived distortion data alone and comparably to an independent
geometry-exact vortex-particle computation of the same case
\citep[$2.6$~dB;][]{SharmaSarradjSchmidt2020}. The scalar rapid-distortion
correction, negligible at these scales, leaves the over-prediction intact
there; semi-analytical distortion models built on the same physics are an
active development \citep{PiccoloJSV2026}, and the present correction
differs in being evaluated from the resolved field rather than modelled. Above $\kappa \approx 1$ the governing physics changes: the blocking
releases eddies smaller than the nose, and the steep computed roll-off is
the established scalar thickness effect
\citep{Gershfeld2004,RogerMoreau2005}. Combining the two --- tensorial for
$\kappa\le1$, scalar beyond --- tracks the full measured spectrum across
$195$--$1599$~Hz to $1.8$~dB (figure~\ref{fig:acoustic_hybrid}).

\subsection{Blind application to published data}
\label{sec:results_blind}

\begin{figure}
  \centering
  \includegraphics[width=0.95\textwidth]{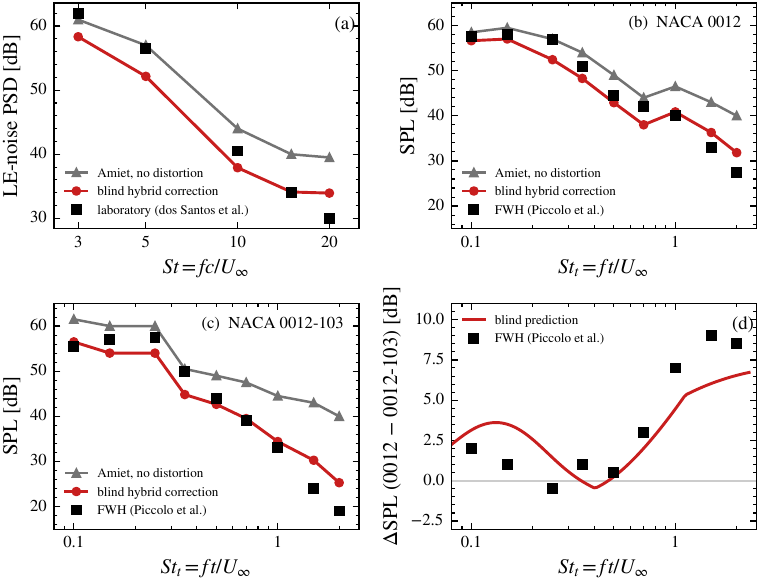}
  \caption{Blind application of the hybrid correction to published data;
  in every panel only the nose radius and flow speed of the target case
  enter. (\textit{a})~laboratory experiment of \citet{dosSantos2023}
  (NACA 0008 in rod turbulence; spectra digitised $\pm1$~dB): flat-plate
  Amiet baseline and the corrected prediction.
  (\textit{b},\textit{c})~lattice-Boltzmann computations of
  \citet{Piccolo2024} for two aerofoils of equal thickness and unequal
  nose radius, against their Ffowcs Williams--Hawkings (FWH) reference.
  (\textit{d})~the pair difference
  $\mathrm{SPL}_{0012}-\mathrm{SPL}_{0012\text{-}103}$, in which the
  flat-plate baseline cancels identically: a baseline-free test of the
  $\kappa=k\,r_{\mathrm{LE}}$ scaling. An interactive \href{https://cocalc.ai/share/7YwVvJLcTMAQ}{JFM Notebook} is available.}
  \label{fig:independent}
\end{figure}

To test transferability we apply the same coefficients, without adjustment, to three independent published data
sets: the coefficients of (\ref{eq:blockfit}) and the computed coherence
stretching are applied unchanged, with only the nose radius and flow
speed of each new case entering (figure~\ref{fig:independent}). The first is a laboratory experiment
in a different facility and a different regime: a NACA~0008 in
rod-generated turbulence \citep{dosSantos2023}, whose integral scale is
$34$ nose radii against our ${\approx}4$. The blind correction reduces
the mean error of the uncorrected flat-plate prediction from $4.1$ to
$2.9$~dB over the leading-edge-noise band, and from $+9.5$ to $+3.9$~dB
at the worst point (figure~\ref{fig:independent}\textit{a}) ---
approaching the accuracy those authors obtain only by re-measuring the
turbulence inside their own facility near the nose. Two of this paper's
central quantities also appear independently in their data: a lateral
coherence stretching of $\times2.8$ (here $\times2.9$), and the finding
that turbulence inputs must be taken at the pre-impact station, two nose
radii ahead of the leading edge. The residual structure is informative
rather than random: the correction over-corrects their very largest
eddies by ${\sim}4$~dB, and the sweep of \S\ref{sec:results_sweep}
explains why --- the low-$\kappa$ blocking anisotropy is constant to within
$\pm10\%$ across the blunt four-digit family
($\Lambda_u/r_{\mathrm{LE}}=2.2$--$8.9$) but vanishes for the
sharp-nosed 66-006 ($\Lambda_u/r_{\mathrm{LE}}=31$), so eddies much
larger than every nose scale begin to escape the blocking. The same
limit recovers the flat plate, where uncorrected thin-aerofoil theory is
known to suffice even in strongly anisotropic turbulence
\citep{HalesJFM2023}.

The second test is numerical and targets the scaling itself. For two
aerofoils of identical thickness whose nose radii differ by $2.8\times$
--- the NACA~0012 / 0012-103 pair computed by an independent group with
a different lattice-Boltzmann solver \citep{Piccolo2024} --- the
flat-plate baseline cancels identically in the difference of the two
spectra, leaving a prediction that depends only on
$\kappa=k\,r_{\mathrm{LE}}$. The blind prediction tracks their computed
pair difference to $1.7$~dB on average
(figure~\ref{fig:independent}\textit{d}), including its change of sign
and its growth to $+8.5$~dB; applied to the individual spectra
(figure~\ref{fig:independent}\textit{b},\textit{c}), it reduces the
mean error from $4.6$ and $8.6$~dB to $2.6$ and $3.2$~dB.
Two of their independent conclusions coincide with values established here:
their fitted distortion length ``almost coincides'' with the nose
radius, and their canonical prediction fails above $St_t=0.35$, which is
$\kappa\approx0.31$ --- the scale at which \S\ref{sec:results_boundary}
found the linear inheritance to collapse.

The third data set takes the correction to its limit, and the limit is
instructive. The laboratory thickness series of \citet{Chaitanya2015},
digitised from its reproduction in \citet{PiccoloJSV2026}, varies
thickness and nose radius \emph{together} --- NACA 0006 to 0018 is a
threefold thickness and ninefold nose-radius change --- at three speeds.
At the energy-containing scales the measured difference between any two
of these aerofoils collapses across the twofold speed change to within
$1$--$2$~dB at matched reduced frequency: the velocity half of the
$\kappa$-scaling, which no other data set here tests. Its magnitude,
however ($9$--$14$~dB for the extreme pair), is more than double the
nose-radius-only prediction: when the thickness changes along with the
nose, the largest eddies respond to the thickness, not the nose ---
the scale-dependent blocking length measured directly by
\citet{dosSantosAIAA2024}, and adopted by an independent wind-turbine
implementation of the same physics \citep{Yalcin2026}. The
same-thickness pair of figure~\ref{fig:independent}\textit{d}, where
that variable is controlled, is precisely where the blind prediction
succeeds. One length does not fit all eddies: the nose radius organises
the distortion of eddies near its own scale, and a second, thickness-set
length takes over for the largest --- the two-scale picture that
\S\ref{sec:generality} develops. At sub-nose scales the measured tails
of the thickest sections approach the facility noise floor, and we draw
no conclusion there.

Table~\ref{tab:validation} collects these tests by workflow. Two entries
are more accurate than the present correction, and both draw on the
target case itself: rapid-distortion theory driven by turbulence measured
\emph{inside the target facility} at the nose \citep{dosSantos2023}, and
a geometry-exact computation of the same case
\citep{SharmaSarradjSchmidt2020}. Among workflows that predict ---
nothing of the target case entering beyond its nose radius and flow
speed --- the present correction is the most accurate on every case
tested here. The nearest competing predictive workflow is the
semi-analytical distortion model of \citet{PiccoloJSV2026}, which
reports ${\approx}3$~dB on a fourth data set from XFOIL-based inputs at
comparable cost; it models geometry effects the present correction does
not (camber, and the thickness boundary above), while its authors note
that the altered spanwise coherence length near the leading edge is the
quantity their modelling captures least convincingly --- precisely the
quantity resolved directly here (figure~\ref{fig:scales}) and carried by
the correction.

\begin{table}
  \begin{center}
  \begin{tabular}{llccc}
    Workflow & Case-specific input &
    \begin{tabular}{@{}c@{}}AWB\\(laboratory)\end{tabular} &
    \begin{tabular}{@{}c@{}}Twente\\(laboratory)\end{tabular} &
    \begin{tabular}{@{}c@{}}Delft pair\\(LBM/FWH)\end{tabular} \\[6pt]
    Amiet, undistorted inputs & ---
      & $+6.6$ & $4.1$ & $4.6$ / $8.6$ \\
    Scalar rapid-distortion filter & $r_{\mathrm{LE}}$
      & slope only & $11$ at $St{=}10$ & --- \\
    Hybrid tensorial correction & $r_{\mathrm{LE}}$, $U_\infty$
      & $1.8$ & $2.9$ & $2.6$ / $3.2$ \\
    RDT, in-situ inputs & hot-wire survey at nose
      & --- & $1.1$ & --- \\
    Vortex-particle computation & resolved geometry
      & $2.6$ & --- & --- \\
  \end{tabular}
  \caption{Accuracy of leading-edge-noise prediction workflows on the
  validation cases of this section, in dB: mean absolute deviation from
  the measured spectrum (AWB, this work; Twente,
  \citealp{dosSantos2023}) or from the FWH reference (Delft pair,
  \citealp{Piccolo2024}) over the leading-edge-noise band of each case.
  Twente and Delft values are evaluated from digitised published spectra
  ($\pm1$~dB); the AWB Amiet entry is the mean over-prediction at the
  energy-containing scales, and the scalar filter corrects the decay
  slope but not the level. Applications of the present correction beyond
  the AWB case are blind: only the target's nose radius and flow speed
  enter. The last two rows require measurements or a computation of the
  target case itself.}
  \label{tab:validation}
  \end{center}
\end{table}

The energy-containing scales, which set the radiated level, are distorted tensorially (blocked in amplitude and stretched in coherence) by the mechanism of \S\ref{sec:mechanism}, while the established scalar correction applies only to the sub-nose scales. What remains is to state what organises these
two scale ranges, and what the results imply beyond this one aerofoil.

\section{Discussion: the two organising scales}
\label{sec:generality}
\label{sec:generality_thickness}

Two distinct scales appear and should be kept separate. The \emph{linear-inheritance} boundary lies
at the large, energy-containing scales: $\gamma^{2}$ collapses already by
$\kappa\approx0.3$, so even eddies several times the nose radius arrive
stripped of their linear image. The \emph{blocking} of the gains is
centred on the nose radius itself: a $-6.8$~dB suppression at
$\kappa\lesssim0.5$, recovering from $\kappa\approx1$. A coarser
computation conflated the two through the under-resolution of its
reference station (mesh-independence check in the supplementary
material); at full resolution they are cleanly distinct, and both
coordinates of the $(\kappa,\vartheta)$ description are computed rather
than estimated ($U_c$ validated upstream, $\vartheta\approx0.68$ on the
stagnation line; supplementary material). Against the three expectations
linear theory posed in \S\ref{sec:results_volume} the outcome is: (i)~confirmed --- the gains are unequal, with the
wall-normal component the most suppressed; (ii)~refined --- the
transition does onset at $\kappa\approx1$, but its completion lies
beyond the resolved band, and the linearity boundary sits at
$\kappa\approx0.3$, below the geometric scale; (iii)~confirmed --- the
trajectory leaves the axisymmetric branch towards $III<0$. The nose-radius half of the scaling is established by the sweep of
\S\ref{sec:results_sweep}; the velocity half is supported, at the
energy-containing scales, by the three-speed laboratory series of
\S\ref{sec:results_prediction}, and its full-spectrum version remains
the falsifiable content of a companion study.

These two scales also connect the present result to the oldest empirical
fact about this noise source: a blunter nose attenuates a wider
high-frequency band, so at matched inflow the thinner section is louder
\citep{Gershfeld2004,Lysak2013,Gill2013,DevenportStaubsGlegg2010}. The
acoustically dominant thickness attenuation operates on the \emph{small}
scales, $\kappa>1$, through the scalar roll-off
$D_{\mathrm{RDT}}=(1+\kappa^{2})^{-5/6}$
\citep{Gershfeld2004,RogerMoreau2005}: a thinner nose shifts the cut-off
upward and is louder at high frequency. The large scales are not passed
unaltered either --- they are transformed tensorially, blocked in
amplitude and stretched in coherence, the two nearly cancelling in the
radiated level. The geometric filter is real and organised by $\kappa$,
but it is not a single high-pass: sub-nose scales are attenuated
scalar-wise while the energy-containing scales are rebuilt by the
mechanism of \S\ref{sec:mechanism} --- a distinction directly relevant to
current prediction efforts for thick sections and wind-turbine inflow
\citep{Piccolo2024,Yalcin2026}.

\section{Conclusions}
\label{sec:conclusions}

We set out to characterise the operator $\mathcal{D}$ that maps the
upstream turbulence tensor onto the pre-impact tensor at an aerofoil
leading edge. Doing so brought a prior question to the fore --- one on which that
operator, and every rapid-distortion correction built upon it, silently
depends: to what extent is the pre-impact turbulence a
\emph{linear image} of the incident field at all? Three congruent,
synchronously sampled surfaces provide the frame in which that question is
well posed and is answered scale by scale.

The answer is a low and receding boundary. Even at the largest resolved
scales little more than a third of the pre-impact spectral energy is
linearly inherited; the inheritance collapses at the energy-containing
scales ($\kappa\approx0.3$, well inside the nose-radius scale), and the
aerofoil-specific part of the loss --- a factor of three below the
control's free-decay baseline --- is localised to the final nose radii.
The same boundary stands in physical space: $III_b$ crosses from prolate
to oblate within the final ${\approx}3\%$ of chord, and the computed
component amplifications leave an exactly evaluated, parameter-free linear
benchmark over the same final approach. Turbulence inputs for predictive
models must be acquired close to the leading edge because the far-upstream
field cannot be carried there as a linear datum --- an empirical rule of
the prediction literature that these results explain.

The mechanism behind both signatures is pressure--strain redistribution.
A traceless identity makes the tensor computable on a single sampling
surface; so evaluated, it drains the strain-amplified wall-normal stress
into the tangential and spanwise components, is absent in the no-aerofoil
control, switches on where the accumulated strain becomes rapid ---
preceding the $III_b$ crossing it produces --- and is corroborated in
sign and location by an independent mean-field budget. The turbulence it
generates is spanwise-rich and large-scale; across a fourteen-fold sweep
of nose radius the anisotropy crossing collapses at
$x\approx-1.5\,r_{\mathrm{LE}}$, the sharpest section marking a bluntness
threshold, while the spanwise-rich generated composition is invariant.
Blocked amplitude ($-6.8$~dB) and stretched spanwise coherence (${\approx}2.9\times$) nearly cancel at the largest scales; the resulting tensorial correction improves agreement with the measured leading-edge noise over the energy-containing band relative to a scalar correction. Applied blind --- nose radius and flow
speed the only inputs --- to published cases from three independent
groups, the correction roughly halves the error of the uncorrected
prediction wherever its regime assumptions hold, and its domain boundary
is fixed by the nose-radius sweep rather than assumed.

The scope of these statements is set by the data that support them: the
largest scales carry the widest confidence intervals, the wall-normal
pressure--strain follows from its traceless identity, the mean-field
budget is read for structure rather than magnitude, and the flow is a
single turbulence state at one Reynolds and Mach number --- a deliberately
clean configuration, chosen so that every link from incident tensor to
radiated sound could be established within one fully resolved flow. That
foundation is precisely what makes the continuations attractive. A
dedicated sweep in free-stream speed and Reynolds number will complete
the $\kappa$-universality test begun by the published multi-velocity
series of \S\ref{sec:results_blind}; loaded and cambered aerofoils,
where the stagnation point migrates and the mean flow itself carries
vorticity, bring the operator into contact with the configurations of practical
flight; and the instrument itself --- congruent surfaces sampled on a common
clock --- transfers to any body that reorganises the turbulence
approaching it. The boundary of the linear description can be measured rather than assumed.

\section*{Acknowledgements}
The authors gratefully acknowledge the High-Performance Computing (HPC)
resources provided by the Computer for Advanced Research in Aerospace (CARA
\& CARO) at DLR, which enabled the numerical simulations presented in this
work.

\section*{Funding}
Financial support from the Exergie project is gratefully acknowledged.

\section*{Declaration of interests}
The authors report no conflict of interest.

\section*{Data availability statement}
The reduced data and analysis underlying the figures accompany this paper as
JFM Notebooks: one self-contained notebook per data figure
(figures~5--11, 13, 14 and 16--24, each bundling the reduced data it plots),
three supplementary-figure notebooks, a step-by-step notebook for
appendix~A, and a documented drop-in implementation of the hybrid correction
(\texttt{len\_distortion\_correction.py}, including its measured validity
envelope). The raw simulation records are available from the corresponding
author on reasonable request.

\section*{Declaration of the use of AI tools}
The authors used a large
language model, Apertus-8B-Instruct-2509 \citep{swissai2025apertus} (Swiss AI
Initiative; hosted on the Blablador service of the J\"ulich Supercomputing
Centre, Helmholtz Association;
accessed July 2026), for English-language grammar and phrasing checks of the
authors' own draft text.

\bibliographystyle{plainnat}
\bibliography{jfm}

\end{document}